\documentclass{jpsj3}
\usepackage{amssymb,amsmath}
\usepackage{graphics}
\usepackage{array}
\usepackage{wrapfig}

\title{Charged Hadron Multiplicity Distribution at Relativistic Heavy Ion Colliders}

\author{Ashwini Kumar, P. K. Srivastava, ~B. K. Singh and C. P. Singh} 

\inst{High Energy Physics Laboratory,~Department of Physics, Banaras Hindu University,~Varanasi 221005, INDIA}

\abst
{ The present article reviews facts and problems concerning charge hadron production in high energy collisions. Main emphasis is laid on the qualitative and quantitative description of general characteristics and properties observed for charged hadrons produced in such high energy collisions. Various features of available experimental data e.g., the variations of charged hadron multiplicity and pseudo-rapidity density with the mass number of colliding nuclei, center-of-mass energies and the collision centrality obtained from heavy-ion collider experiments are interpreted in the context of various theoretical concepts and their implications. Finally, several important scaling features observed in the measurements mainly at RHIC and LHC experiments are highlighted in the view of these models to draw some insight regarding the particle production mechanism in heavy-ion collisions.     }

\kword{ Nucleus-nucleus interactions, multiplicity, pseudo-rapidity density, collision energy, centrality.}
\begin{document}



\maketitle

\section{Introduction}
Quantum chromodynamics (QCD), the basic theory which describes the interactions of quarks and gluons is a firmly established microscopic theory in high energy collision physics. Heavy-ion collision experiments provide us a unique opportunity to test the predictions of QCD and simultaneously to understand the two facets of high energy collision process: hard process (i.e, the small cross-section physics) and soft process (i.e, the large cross-section physics)~\cite{[gribov],[levin],[dremin]}. Nuclear collisions at very high energies such as collidier energies enable us to study the novel regime of QCD, where parton densities are high and the strong coupling constant between the partons is small which further decreases as the distance between the partons decreases. The parton densities in the initial stage of the collision can be related to the density of charged hadrons produced in the final state. With the increase in collision energy, the role of hard process (minijet and jet production) in final state particle production rapidly increases and offers a unique opportunity to investigate the interplay between various effects. In this scenario, the perturbative QCD (pQCD) lends a good basis for high energy dynamics and has achieved significant success in describing hard processes occuring in high energy collisions such as scaling violation in deep inelastic scattering DIS~\cite{[castorina]}, hadronic-jet production in $e^{+} e^{-}$ annihilation~\cite{[hanson],[berger]}, large-pt-jet production in hadron-hadron collisions~\cite{[aad],[bern],[nagy1],[ellis],[giele]}, etc. On the other hand, in soft processes such as hadron production with sufficiently small transverse momentum in hadronic and nuclear collisions, the interactions become so strong that the perturbative QCD (pQCD) does not remain applicable any more. Thus, there is no workable theory yet for non-perturbative QCD regime which can successfully describe these soft processes.  Due to inapplicability of pQCD in this regime, experimental input-based phenomenological models are proven to be an alternative tool to increase our knowledge on the  property of the basic dynamics involved in such collision processes. Furthermore, these soft hadrons which decouple from the collision zone in the late hadronic freezeout stage of the evolution are quite useful in providing the indirect information about the early stage of the collision. Several experimental information on the multiparticle production in lepton-hadron, hadron-hadron, hadron-nucleus and nucleus-nucleus have been accumulated in recent past over a wide range of energy. In this context, the bulk features of multiparticle production such as the average charged particle multiplicity and particle densities are of fundamental interest as their variations with the collision energy, impact parameter and the collision geometry are very sensitive to the underlying mechanism involved in the nuclear collisons. These can also throw some more light in providing insight on the partonic structure of the colliding nuclei. In order to understand the available experimental data, a lot of efforts have been put forward in terms of theoretical and phenomenological models. However, the absence of any well established alternative, the existing  problem of the production mechanism of charged hadrons continues to facilitate proliferation of various models. The development in this direction is still in a state of flux for describing the same physical phenomenon using different concepts and modes of operation.  Most of these theoretical models are based on the geometrical, hydrodynamical and statistical approaches.  However, the diverse nature of the experimental data poses a major challenge before physics community to uncover any systematics or scaling relations which are common to all type of reactions. Thus, a search for universal mechanism of charged hadron production common to hadron-hadron, hadron-nucleus and nucleus-nucleus interactions is still continued and  needs a profound effort to draw any firm conclusion. Also, the complicated process of many-body interactions occuring in these collision processes is still quite difficult to make a clear understanding of the phenomena by analyzing the experimental data on the multiparticle production in the final state. In this regard, ongoing efforts for the extensive analysis of the experimental data available on  charged hadron production in the view of some successsful phenomenological models can provide us a much needed insight in developing a better understanding of the mechanism involved in the particle production. Moreover, these can also be useful in revealing the properties of the nuclear matter formed at extreme conditions of energy and matter densities.  

In this review, we attempt to give a succint description of most of the progress made in this field till date even it is not so easy for us. Furthermore, we believe that the references mentioned in this review will surely guide the readers  but we can never claim that they are complete. We apologize to those authors whose valuable contributions in this field have not been properly mentioned.     

The structure of this paper is framed in the following manner: At first, In section 2, we start with a brief description of different models used for the study of charged hadron productions in this review in a systematic manner. In section 3, the experimental results on  charged hadron production at collider energies are presented along with the comparison of different model results. Further in section 4, we will provide some scaling relations for charged hadrons production and evaluate them on the basis of their universality in different collisions.

\section{Model Descriptions}
\subsection{\bf{Wounded nucleon model}}
In 1958, Glauber presented his first collection of various papers and unpublished work~\cite{[glauber]}.  Before that, there were no systematic calculations for treating the many-body nuclear system as either a projectile or target. Glauber's work put the quantum theory of collisions of composite objects on a firm basis. In 1969 Czyz and Maximon~\cite{[czyz]} applied the Glauber's theory in its most complete form for proton-nucleus and nucleus-nucleus collisions. By using Glauber's theory, finally A. Bialas et al.~\cite{[bialas]} first proposed the wounded nucleon model which was based on the basic assumption that the inelastic collisions of two nuclei can be described as an incoherent composition of the collisions of individual nucleons of the colliding nuclei. In this approach, the collective effects which may occur in nuclei were neglected. According to their assumption, in nucleon-nucleus collisions, a fundamental role is played by the mean number of collisions $\nu$ suffered by the incident nucleon with the nucleons in the target nucleus~\cite{[gottfried]}. Similarly, nucleus-nucleus collision is also described in terms of the number of wounded nucleons ($w$).
For the nucleon-nucleus collisions, there is a simple relation between $\nu$ and $w$
\begin{equation}
w = \nu + 1
\end{equation}
but no such type of relation exist for the nucleus-nucleus collisions.
Motivated by the data available on nucleon-nucleus ($N-A$) interactions~\cite{[florian]}, the average multiplicity  follows approximately the formula
\begin{equation}
\bar{n}_A=\frac{1}{2}(\bar{\nu}+1)=\frac{1}{2}\bar{w}\bar{n}_H,
\end{equation}  
where $\bar{n}_{H}$ is the average multiplicity in nucleon-nucleon collisions.
For nucleus-nucleus (A-B) collisions generalization of this picture implies that the average multiplicity  is
\begin{equation}
\bar{n}_{AB}=\frac{1}{2}\bar{w}_{AB}\bar{n}_H,
\end{equation}
where the number of wounded nucleons (participants) in the collision of $A$ and $B$ is the sum of wounded nucleons in the nucleus $A$ and the nucleus $B$ i.e., $\bar{w}_{AB} = \bar{w}_{A}+\bar{w}_{B}$ with $\bar{w}_{A} = A\frac{\sigma_{B}}{\sigma_{AB}}$ and  $\bar{w}_{B} = B\frac{\sigma_{A}}{\sigma_{AB}}$. The extension of Glauber model was used to describe elastic, quasi-elastic, the total cross sections~\cite{[czyz],[faldt],[barshay],[fishbane],[franco],[bialas1]}.
In the wounded nucleon model, the basic entity is the nucleon-nucleon collision profile $P(b)$ defined as the probability of inelastic nucleon-nucleon collision at impact parameter $b$. Once the probability of a given nucleon-nucleon interaction is known, the probability of having n such interactions in collision of nuclei $A$ and $B$ is given as~\cite{[miller]}:
\begin{equation}
P(n,b)=\left( \begin{array}{c} AB \\ n \end{array} \right)\left[T_{AB}(b){\sigma_{inel}^{NN}}\right]^{n}\left[1-T_{AB}(b){\sigma_{inel}^{NN}}\right]^{AB-n},
\end{equation} 
where $T_{AB}(b)$ is thickness function and the $\sigma_{inel}^{NN}$ is the inelastic nucleon-nucleon cross-section. The total cross-section is given by:
\begin{equation}
\sigma_{inel}^{AB} = \int_{0}^{\infty}{2\pi bdb \left\{1-[1-T_{AB}(b)\sigma_{inel}^{NN}]^{AB}\right\}}.
\end{equation}
The total number of nucleon-nucleon collisions is given by:
\begin{equation}
N_{coll}(b) = \sum_{n=1}^{AB}nP(n,b)=ABT_{AB}(b)\sigma_{inel}^{NN}.
\end{equation} 
The number of participants ( wounded nucleons) at a given impact parameter is given by:
\begin{equation}
N_{part}(b) = A\int T_{A}(s)\left\{1-[1-T_{B}(s-b)\sigma_{inel}^{NN}]^B\right\}d^{2}s + B\int T_{B}(s)\left\{1-[1-T_{A}(s)\sigma_{inel}^{NN}]^B\right\}d^{2}s.
\end{equation}
with $T_{A(B)}(s) = \int\rho_{A(B)}(s,z_{A(B)})dz_{A(B)}$ is the probability per unit transverse area of a given nucleon being located in the target flux tube of A or B and $\rho_{A(B)}(s,z_{A(B)})$ is the probability per unit volume, normalized to uniy for finding the nucleon at location $(s,z_{A(B)})$. 

The rapidity density of particles in nucleus-nucleus ($A-B$) collision is given by~\cite{[bialas2]}: 
\begin{equation}
\frac{dN_{AB}}{dy}=w_{A}F_{A}(y) +w_{B}F_{B}(y) = \frac{1}{2}(w_{A}+w_{B})[F_{A}(y)+F_{B}(y)]+ \frac{1}{2}(w_{A}-w_{B})[F_{A}(y)-F_{B}(y)],                
\end{equation}
where $y$ is the rapidity in c.m. system, $F_{A}(y)$ and $F_{B}(y)$ are the contribution from a single wounded nucleon in $A$ and $B$. Thus: 
\begin{equation}
F_{B}(y) = F_{A}(-y)                
\end{equation}
The model gives a good description of the data, with the condition (9) being well satisfied, except at rapidities close to the maximal values~\cite{[bialas2]}.

It can be seen from Eq.(8) that for nucleon-nucleon collision, one have:
\begin{equation}
\frac{dN_{NN}}{dy}=F_{A}(y) +F_{B}(y)= F_{N}(y) +F_{N}(-y)               
\end{equation}
and thus for the ratio
\begin{equation}
R_{AB}(y)\equiv \frac{dN_{AB}/dy}{dN_{NN}/dy}               
\end{equation}
one obtains 
\begin{equation}
R_{AB}(y)=\frac{1}{2}(w_{A}+w_{B})+\frac{1}{2}(w_{A}-w_{B})\frac{F_{A}(y)+F_{B}(y)}{F_{A}(y)-F_{B}(y)}               
\end{equation}
consequently, one have at $y$=0,
\begin{equation}
R_{AB}(y=0)=\frac{1}{2}(w_{A}+w_{B}),             
\end{equation} 
which implies that the value of the ratio $R_{AB}$ at mid-rapidity is fully determined by the number of wounded nucleons and independent of the function $F(y)$.
Unlike the $N_{part}$ scaling observed in charged hadron multiplicity, pseudo-rapidity density at mid-rapidity does not scale linearly with $N_{part}/2$~\cite{[miller]}. It was conjectured~\cite{[blaizot],[lindfors],[lindfors1],[xwong],[geiger]} that at sufficiently high energy the particle production in nucleus-nucleus collisions will be dominated by hard processes. However, the gross features of particle production at CERN SPS energies were found to be approximately consistent~\cite{[jones]} with the $N_{part}$ scaling  as accomodated by the wounded nucleon model. A better agreement with the data is found in two-component model for estimating the pseudo-rapidity density in wounded nucleon model as shown by Ref.~\cite{[kharzeev]}. In A-A collisions hadron production from the two processes scales as N$_{part}$/2 (number of nucleon participant pairs) and N$_{bin}$ (number of binary N-N collisions), respectively~\cite{[kharzeev]}. According to this assumption the pseudo-rapidity density of charged hadrons is given as:
\begin{equation}
\frac{dN}{d\eta}=n_{pp}[(1-x)N_{part}+xN_{coll}],                
\end{equation}  
where $x$ quantifies the relative contributions of two components arising from hard and soft processes. The fraction $x$ corresponds to the contribution from hard processes and the remaining fraction ($1-x$) describes the contributions arising from the soft processes. 
\subsection{\bf{Wounded quark Model}}
Charged hadron production by using the concept of the constituent or wounded quarks has been widely used for many years~\cite{[anisovich],[v.anisovich],[hwa],[hwa1]}. In the wounded quark picture, nucleus-nucleus collisions are effectively described in terms of the effective number of constituent quarks participating in the collision process along with the effective number of collisions suffered by each of them. Recently, the idea of wounded quarks was resurrected by S. Eremin et al.~\cite{[eremin]} in which they modified the overlap function by increasing the nucleon density three times and introduced one more parameter the quark-quark interaction cross-section, which reproduced the data well. They have further shown that the charged hadron density at mid-rapidity  can be described well by the wounded quark model. This problem was further investigated in several papers~\cite{[netrakanti],[bhaskar],[nouicer],[sarkisyan],[sakharov]} representing the analysis of various spectra, SPS data, total multiplicities and the energy deposition. De et al.~\cite{[bhaskar]} have shown that the data on $K^{\pm}$ and $p/\bar{p}$ favors the $N_{q-part}$ scaling over the $N_{part}$ scaling whereas the pions do not agree well with such scaling law. Recently, we proposed a wounded quark model~\cite{[ashwini]} which is primarily based on the previous work by C. P. Singh et al.~\cite{[c.p.singh],[c.p.singh1],[c.p.singh2]}. In this picuture, during the collision, a gluon is exchanged between a quark of projectile or first nucleus and a quark belonging to target or other collliding nucleus. The resulting colour force is then somewhat stretched between them and other constituent quarks because they try to restore the colour singlet behaviour. When two quarks separate, the colour force builds up between them and the energy in the colour-field increases, the colour tubes thus formed finally break-up into new hadrons and/or quark-antiquark pairs. We consider a multiple collision scheme in which a valence quark of the incident nucleon suffers one or more inelastic collisions with the quarks of target nucleons. In a nucleon-nucleon collision only one valence quark of each nucleon (i.e, target and projectile) interacts while other quarks remain as spectators~\cite{[sakharov]}. Only a part of the entire nucleon energy is spent for secondary production at mid-rapidity. The other spectator quarks are responsible for  forming  hadrons in the nucleon fragmentation region. In the case of nucleus-nucleus collisions,  more than one quark per nucleon interacts and each quark suffers more than one collision due to a large nuclear size since large travel path inside the nucleus becomes available. If we search a universal mechanism of charged particle production in the hadron-hadron, hadron-nucleus and nucleus-nucleus collisions, it must be driven by the available amount of energy required for the secondary production, and it also depends on the mean number of participant quarks. The main ingredients of our model are taken from the paper by Singh et. al.~\cite{[c.p.singh],[c.p.singh1],[c.p.singh2]}. Charged hadrons produced from $A-A$ collisions are assumed to result from a somewhat unified production mechanism common to $p-p$ collisions at various energies.

Based on the experimental findings by PHOBOS Collaboration, recently Jeon and collaborators have shown that the total multiplicity obtained at RHIC can be bounded by a cubic logarithmic term in energy~\cite{[jeon]}. Therefore, we propose here a new parameterization involving a cubic logarithmic term  so that the entire $p-p$ experimental data starting from low energies (i.e. from 6.15 GeV) upto the highest LHC energy (i.e. 7 TeV)  can suitably be described as~\cite{[ashwini]}:
\begin{equation}
 <n_{ch}> _{hp}=(a'+b' ln \sqrt{s_{a}}+c'ln^{2} \sqrt{s_{a}}+d'ln^{3} \sqrt{s_{a}})-\alpha.
\end{equation}
In Eq. (15), $\alpha$ is the leading particle effect and $\sqrt{s_{a}}$ is the available center-of-mass energy (i.e., $\sqrt{s_{a}}=\sqrt{s}-m_{B}-m_{T}$, where $m_{B}$ is the mass of the projectile and $m_{T}$ the mass of the target nucleon, respectively), $a'$, $b'$, $c'$ and $d'$ are constants derived from the best fit to the $p-p$ data and the value of $\alpha$ is taken here as 0.85~\cite{[ashwini]}.\\
 We can extrapolate the validity of this parametrization further for the produced charged particles in hadron-nucleus interactions by considering multiple collisions suffered by the quarks of hadrons in the nucleus. The number of constituent quarks which participate in hadron-nucleus (h-A) collisions share the total available  center-of-mass energy $\sqrt{s_{hA}}$ and thus the energy available for each interacting quark becomes $\sqrt{s_{hA}}/N_q^{hA}$, where $N_q^{hA}$ is the mean number of constituent quarks in h-A collisions. The total available squared center-of-mass energy  $s_{hA}$ in h-A collisions is related to $s_a$  as ${s_{hA}} = {\nu_q^{hA}s_a}$  with $\nu_{q}^{hA}$ as the mean number of inelastic collisions of quarks with target nucleus of atomic mass A. Within the framework of the Additive Quark Model~\cite{[anisovich],[v.anisovich],[anisovich1],[a.bialas1],[devidenko],[kobrinsky],[nyiri]}, the mean number of collisions in hadron-nucleus interactions is defined as the ability of constituent quarks in the projectile hadron to interact repeatedly inside a nucleus. 
Finally, the expression for average charged hadron multiplicity in h-A collisions is~\cite{[c.p.singh],[c.p.singh1],[c.p.singh2],[ashwini]}:
\begin{eqnarray}
 <n_{ch}> _{hA}=N_{q}^{hA}&\Biggl[&a'+b' ln \left(\frac{\sqrt{s_{hA}}}{N_{q}^{hA}}\right)+c'ln^{2}\left(\frac{\sqrt{s_{hA}}}{N_{q}^{hA}}\right) 
+d'ln^{3}\left(\frac{\sqrt{s_{hA}}}{N_{q}^{hA}}\right)\Biggr] - \alpha.
\end{eqnarray}
The generalization of the above picture for the case of nucleus-nucleus collisions is finally achieved as follows~\cite{[ashwini]}:
\begin{eqnarray}
<n_{ch}> _{AB}=N_{q}^{AB}&\Biggl[&a'+b' ln \left(\frac{\sqrt{s_{AB}}}{N_{q}^{AB}}\right)+c'ln^{2}\left(\frac{\sqrt{s_{AB}}}{N_{q}^{AB}}\right) 
+d'ln^{3}\left(\frac{\sqrt{s_{AB}}}{N_{q}^{AB}}\right) \Biggr],
\end{eqnarray}

The parametrization in Eq. (17) thus relates nucleus-nucleus collisions to hadron-nucleus and further to hadron-proton collisions and the values of the parameters $a', b', c'$, and $d'$ remain unaltered which shows its universality for all these processess.

In creating quark gluon plasma (QGP), greater emphasis is laid on the central or head-on collisions of two nuclei. The mean multiplicity  in central collisions can straight forwardly be generalized~\cite{[ashwini]} from Eq. (17) as :
\begin{eqnarray}
<n_{ch}>^{central}_{AB}=A&\Biggl[&a'+b'ln(\nu_{q}^{AB}s_{a})^{1/2}+c'ln^{2}(\nu_{q}^{AB}s_{a})^{1/2}
+d'ln^{3}(\nu_{q}^{AB}s_{a})^{1/2}\Biggr].
\end{eqnarray}
 The pseudo-rapidity distribution of charged particles is another important quantity in the studies of particle production mechanism from high energy $h-h$ and $A-B$ collisions, which, however, is not yet understood properly. It has been pointed out that $(dn_{ch}/d\eta)$ can be used to get the information on the temperature ($T$) as well as energy density ($\rho$) of the QGP ~\cite{[x.n.wang],[sousa],[eskola]}. For the pseudo-rapidity density of charged hadrons, we first fit the experimental data of $(dn_{ch}/d\eta)^{pp}_{\eta=0}$ for collision-energy ranging from  a low energy to very high energy. One should use the  parameterization upto squared logarithmic term in accordance with  Ref.~\cite{[jeon]}.
Hence,  using a parameterization for central rapidity density as:
\begin{eqnarray}
 <(dn_{ch}/d\eta)^{pp}_{\eta=0}> &=&(a_{1}^{'}+b_{1}^{'} ln \sqrt{s_{a}}+c_{1}^{'}ln^{2} \sqrt{s_{a}})-{\alpha_1}',\nonumber \\
\end{eqnarray}
we obtain the values of the parameters $a_{1}' = 1.24$, $b_{1}' = 0.18$, and $c_{1}' = 0.044$ from the reasonable fit to the p-p data~\cite{[ashwini]}.

Earlier many authors have attempted to calculate the pseudo-rapidity density of charged hadrons in a two component model of parton fragmentation~\cite{[trainor],[a.trainor]}. Its physical interpretation is based on a simple model of hadron production: longitudinal projectile nucleon dissociation (soft) and transverse large-angle scattered parton fragmentation (hard). However, this assumption which is based on a nucleon-nucleon collision in the Glauber model is crude and it looks unrealistic to relate participating nucleons and nucleon-nucleon binary collisions to soft and hard components at the partonic level. Here we modify the two component model of pseudo-rapidity distributions in A-A collisions as given in Eq. (14) for the wounded quark scenario and assume that the hard component which  basically arises due to multiple parton interactions~\cite{[t.a.trainor]} scales with the number of quark-quark collisions (i.e. $N_{q}^{AB}\nu_{q}^{AB}$) and soft component scales with the number of participating quarks (i.e. $N_{q}^{AB}$). Thus, the expression for $(dn_{ch}/d\eta)^{AB}_{\eta=0}$ in A-B collisions can be parameterized in terms of p-p rapidity density as follows~\cite{[ashwini]}:
\begin{equation}
\left(\frac{dn_{ch}}{d\eta}\right)^{AB}_{\eta=0}=\left(\frac{dn_{ch}}{d\eta}\right)^{pp}_{\eta=0}\left[\left(1-x\right)N_{q}^{AB}+ x N_{q}^{AB}\nu_{q}^{AB}\right],
\end{equation}

In order to incorporate $\eta$ dependence in central A-B collisions, we further extend the model by using the functional form: 
\begin{equation}
\left(\frac{dn_{ch}}{d{\eta}}\right)^{AB} = 2\left(\frac{dn_{ch}}{d\eta}\right)^{AB}_{\eta = 0} \frac{\sqrt{1- \frac{1}{(\beta cosh {\eta})^2}}}{\gamma + exp (\eta^2/2{\sigma}^2)},
\end{equation}
where $\beta, \gamma$ and $\sigma$ are fitting parameters and $({dn_{ch}}/{d\eta})^{AB}_{\eta = 0}$ is the central pseudo-rapidity density in  A-B collisions obtained from Eq. (20). 

\subsection{\bf{Dual Parton Model}}
Dual parton model was introduced at Orsay in 1979, by incorporating partonic ideas into the Dual topological unitarization (DTU) scheme~\cite{[dpm],[dpmm1],[dpmm2],[dpmm3],[dpmm4]}.Dual parton model (DPM) and Quark gluon string model (QGSM) are multiple-scattering models in which each inelastic collision results from the superposition of two strings and the weights of the various multiple-scattering contributions are represented by a perturbative Reggeon field theory. One assumes a Poisson distribution for each string for fixed values of the string ends. The broadening of distribution  arises due to the fluctuations in the number of strings and to the fluctuation of the string ends. When the effect of the fluctuations of the string ends is negligibly small, then DPM reduces to an ordinary multiple scattering model with identical multiplicities in each individual scatterings~\cite{[dpm1]}. The inclusive spectra for charged particle multiplicity in $p-p$ collisions is given as follows~\cite{[review]}:\begin{equation}
\frac{dN^{pp}}{dy}= \frac{1}{\sigma_{in}}\frac{d\sigma^{pp}}{dy}=\frac{1}{\sum{\sigma_k}}\sum{\sigma_k}[N_{k}^{qq-q}(s,y)+N_{k}^{q-qq}(s,y)+N_{k}^{q-\bar{q}}(s,y)],
\end{equation} 
when all string contributions are identical means all individual scattering are same then this expression reduces to the simple expression
\begin{equation}
\frac{dN^{pp}}{d\eta}= \frac{1}{\sigma_{ND}^{pp}}\frac{d{\sigma}^{pp}}{d\eta}=\frac{1}{\sum_{k\ge 1} {\sigma_k}}\sum_{k\ge 1} \sigma_{k} k \frac{dN_{0}^{pp}}{d\eta}=\langle k \rangle \frac{dN_{0}^{pp}}{d\eta},
\end{equation} 
here $\langle k \rangle$is the average number of inelastic collisions and $\frac{dN_{0}^{pp}}{d\eta}$ the charged multiplicity per unit pseudo-rapidity in an individual $p-p$ collision.

To calculate the weights $\sigma_k$ for the occurence of $k$ inelastic collisions, a quasi-eikonal model has been used. The $\sigma_k$ is given as follows~\cite{[dpm1]}:
\begin{equation}
\sigma_k(\xi)=\frac{\sigma_p}{k Z}[1-exp(-Z)\sum_{i=0}^{k-1}\frac{Z^i}{i!}](k\ge1),
\end{equation} 
Here $\xi = ln(s/s_o)$, $\sigma_p=8\pi\gamma_p exp(\Delta \xi)$ and $Z = {2 C_{E}\gamma_p exp(\Delta \xi)}/(R^2 + \alpha_{p}'\xi)$.

In Eq. (23), $\sigma_p$ is the Born term given by Pomeron exchange with intercept $\alpha_p(0) = 1+\Delta$. According to a well known identity which is known as AGK cancellation~\cite{[agk]}, all multiple scattering contributions vanishes identically in the single particle inclusive distribution $d\sigma/d\eta$ and only the Born term ($\sigma_p$) contribution is left. The parameter $R^2$ and $\alpha_{p}'$ control the t-dependence of the elastic peak and $C_E$ contain the contribution of diffractive intermediate states. The total $p-p$ cross-section in this prescription is~\cite{[capella2013]}:
\begin{equation}
\sigma_{tot}(s)= \sigma_{p} f(z/2), f(z) = \sum_{l=1}^{\infty}\frac{(-z)^{l-1}}{ll!}
\end{equation}  
Thus one can calculate the pseudo-rapidity distribution of charged hadron from Eq.(22) and using expectation value of $\langle k \rangle$ as follows:
\begin{equation}
\langle k \rangle = \frac{\sum_{k\ge1}k\sigma_k}{\sum_{k\ge1}\sigma_k}=\frac{\sigma_p}{\sigma_{ND}^{pp}}
\end{equation}  
Further, the cross-section for $\nu$ inelastic collisions in $pA$ scattering after considering the AGK cancellation is as follows~\cite{[dpm1]}:
\begin{equation}
\sigma_{\nu}^{pA}(b)=\left( \begin{array}{c} A \\ \nu  \end{array} \right)(\sigma_{pp}T_{A}(b))^{\nu}(1-\sigma_{pp}T_{A}(b))^{A-\nu}.
\end{equation} 
Here $T_{A}(b)$ represents the nucleon profile function as defined in the wounded nucleon model. The first factor on the R.H.S. of Eq.(27) yields the number of ways  $\nu$ interacting nucleons can be chosen out of $A$. The second factor is the probability that $\nu$ nucleons interact at fixed parameter $b$. The third factor is the probability for no interaction of the remaining $A-\nu$ nucleons. On the contrary the multiplicity for $p-A$ scattering in DPM is given as follows:
\begin{equation}
\frac{dN^{pA}}{d\eta}=\frac{1}{\sigma_{pA}}\sum_{\nu=1}^{A}\sigma_{nu}^{pA}\frac{dN^{pp}}{d\eta},
\end{equation}
with $\bar{\nu} = A\sigma_{pp}/\sigma_{pA}$. Thus $\frac{dN^{pA}}{d\eta}$ scales with the number of binary collisions. Further, the AGK cancellation implies that at mid-rapidity $d\sigma^{AA}/d\eta = A^2 d\sigma^{pp}/d\eta$, which implies 
\begin{equation}
\frac{dN^{AA}}{d\eta}=A^2 \frac{\sigma_{pp}^{ND}}{\sigma_{AA}}\frac{dN^{pp}}{d\eta}=n_{coll}\frac{dN^{pp}}{d\eta},
\end{equation}
As a function of the impact parameter, the average number of binary nucleon-nucleon collisions $n_{coll}(b)$ can be expressed as follows~\cite{[dpm1]}:
\begin{equation}
n_{coll}(b)=A^2\frac{\sigma_{pp}^{ND}}{\sigma_{AA}(b)}T_{AA}(b)
\end{equation}
Capella and Ferreiro~\cite{[dpm1]} authors have also incorporated the corrections arising due to shadowing effects by including the contribution of triple Pomeron graphs. The suppression in multiplicity from shadowing in $A-A$ collision for a particle produced at mid-rapidity can be obtained by replacing the nuclear profile function $T_{AA}(b)=\int d^{2}sT_{A}(s)T_{A}(b-s)$: 
\begin{equation}
S^{sh}(b,s)=\int d^{2}s \frac{T_{A}(s)}{1+AF(y=0)T_{A}(s)}\times \frac{T_{A}(b-s)}{1+AF(y=0)T_{A}(b-s)},
\end{equation} 
where $F(y=0) = {c[exp(\Delta y_{max})-exp(\Delta y_{max})]}/{\Delta}$. Further in Eq. (29) and Eq. (30), $\sigma_{AA}(b)$  can be obtained as follows:
\begin{equation}
\sigma_{AA}(b)=1-exp[-\sigma_{pp}^{ND}A^2T_{AA}(b)]
\end{equation} 

\subsection{\bf{The Color Glass Condensate Approach}}
Color Glass Condensate~\cite{[mclarren],[mclarren1],[mclarren2],[mclarren3],[mclarren4],[kovchegov],[balitsky],[jalilian],[jalilian1],[jalilian2],[jalilian3],[jalilian4],[jalilian5],[jalilian6],[iancu],[iancu1],[iancu2]} (CGC) is an effective theory that describes the gluon content of a high energy hadron or nucleus in the saturation regime. At high energies, the nuclei gets contracted and the gluon density increases inside the hadron wave functions and at small $x$, the gluon density is very large in comparison to all other parton species (the valence quarks) and the sea quarks are suppressed by the coupling $\alpha_s$, since they can be produced from the gluons by the splitting $g\to q\bar{q}$. System that evolve slowly compared to natural time scales are generally glasses. The word Color is because CGC is composed of colored gluons. The word Glass is because the classical gluon field is produced by fast moving static sources. The distribution of these sources is real. System that evolves slowly compared to natural time scales are generally glasses. The word condensate means the gluon distribution has maximal phase space density for momentum modes and the strong gluon fields are self generated by the hadron. This effective theory approximates the description of the fast partons in the wave function of a hadron. This framework has been applied in a range of experiments e.g., from DIS to proton-proton, proton-nucleus and nucleus-nucleus collisions. One of the early success of the CGC was the description of multiplicity distributions in DIS experiments~\cite{[eskola]}. When the nucleus is boosted to a large momentum-then due to Lorentz contraction in the transverse plane of nuclei, partons have to live on thin sheet in the transverse  plane. Each parton occupies the transverse area $\pi/Q^2$ and can be probed with the cross section $\sigma \sim \alpha_{s}(Q^2)\pi/Q^2$. On the other side the total transverse area of the nucleus is $S_A \sim \pi R_{A}^{2}$. Therefore if the number of partons exceeds
\begin{equation}
N_A \sim S_A/\sigma \sim \frac{1}{\alpha_{s}(Q^2)}Q^2 {R_A}^2
\end{equation}
then they start to overlap in the transverse plane and start interacting with each other which prevents further growth of parton densities. At this situation the transverse momenta of the partons are of the order of ${Q_s}^2 \sim \alpha_{s}(Q_{s}^2)\frac{N_A}{R_{A}^2} \sim A^{1/3}$, which is called as ``saturation scale''~\cite{[kharzeev]}.

The multiplicity of the produced partons should be proportional to~\cite{[kharzeev]}:
\begin{equation}
N_s \sim \frac{1}{\alpha_{s}(Q_{s}^2)}Q_{s}^2 R_{A}^2 \sim N_A \sim A
\end{equation}
In the first approximation, the multiplicity in this high density regime scales  with the number of participants. However, there is an important logarithmic correction to this from the evolution of parton structure functions with $Q_{s}^2$. The coefficient of proportionality is given as follows~\cite{[kharzeev]}:
\begin{equation}
Q_{s}^2 = \frac{8 {\pi}^2 N_c}{N_{c}^2 -1}\alpha_{s}(Q_{s}^2)xG(x,Q_{s}^2)\frac{\rho_{part}}{2},
\end{equation} 
where $N_c$=3 is the number of colour, $xG(x,Q_{s}^2)$ is the gluon structure function in nucleus and $\rho_{part}$ is the density of participants in the transverse plane.

Number of produced partons is given as follows~\cite{[kharzeev]}:
\begin{equation}
\frac{d^2{N}}{d^2{b}d\eta} = c \frac{N_{c}^2-1}{4{\pi}^2 N_c}\frac{1}{\alpha_s}Q_{s}^2,
\end{equation} 
where c is the ``parton liberation'' coefficient accounting for the transformation of virtual partons in the initial state to the on-shell partons in the final state.
Integrating over transverse coordinate and using Eq.(35), one can obtain:~\cite{[kharzeev]}
\begin{equation}
\frac{dN}{d\eta} = c N_{part}xG(x,Q_{s}^2),
\end{equation} 
Here c is assumed to be close to unity in the context of local ``parton hadron duality'' hypothesis~\cite{[dokshitzer]} according to which at $p_{\perp}\sim Q_s$ , the distribution of produced hadrons mirrors that of the produced gluons.
Using Eq. (35) and (37), one can evaluate the centrality dependence as follows~\cite{[kharzeev]}:
\begin{equation}
\frac{2}{N_{part}} \simeq 0.82 ln\left(\frac{Q_{s}^2}{\Lambda_{QCD}^2}\right),\end{equation} 
Here $\Lambda_{QCD}$ is QCD scale parameter.

The rapidity density can be evaluated in CGC Effective Field Theory (EFT) by using the following expression~\cite{[kharz]}:
\begin{equation}
\frac{dN}{dy} = \frac{1}{\sigma_{AA}}\int d^2{p_t}\left(E\frac{d\sigma}{d^3{p}}\right), 
\end{equation} 
where $\sigma_{AA}$ is the inelastic cross-section of nucleus-nucleus interaction. Further, the differential cross-section of gluon production in a $A-A$ collision is written as:~\cite{[kharz]}
\begin{equation}
E\frac{d\sigma}{d^3{p}} = \frac{4\pi N_c}{N_{c}^2-1}\frac{1}{p_{t}^2}\int d^2{k_t}\alpha_{s}\phi_{A}(x_1,k_{t}^2)\phi_{A}(x_2,(p-k)_{t}^2), 
\end{equation} 
where $x_{1,2}=(p_t/\sqrt{s})exp(\pm\eta)$ with $\eta$ the pseudo-rapidity of the produced gluon, $\alpha_s$ is the running coupling evaluated at the scale $Q^2$ = max$\left\{k_{t}^2,(p-k)_{t}^2\right\}$. $\phi_{A}$ is the unintegrated gluon distribution which describes the probability to find a gluon with a given $x$ and transverse momentum $k_t$ inside the nucleus. When $p_{t}^2 > Q_{s}^2$, $\phi_{A}$ corresponds to the bremsstrahlung radiation spectrum and can be expressed as:
\begin{equation}
\phi_{A}(x,k_{t}^2) \sim \frac{\alpha_s}{\pi}\frac{1}{k_{t}^2} 
\end{equation}
In the saturation regime,
\begin{equation}
\phi_{A}(x,k_{t}^2) \sim \frac{S_{A}}{\alpha_s}, k_{t}^2 \le Q_s, 
\end{equation}
where $S_A$ is the nuclear overlap area. 

Since the rapidity ($y$) and Bjorken variable $x$ are related by $ln{\frac{1}{x}}=y$, the $x$-dependence of the gluon structure function translates into the following rapidity dependence of the saturation scale factor $Q_{s}^2$
\begin{equation}
Q_{s}^2(s;\pm y) = Q_{s}^2(s;\pm y=0)exp(\pm \lambda y). 
\end{equation}
Integrating over transverse momentum in (39) and (40), the rapidity distribution comes as follows~\cite{[kharz]}:
\begin{equation}
\frac{dN}{dy} = const \times S_A Q_{s,min}^2 ln\left(\frac{Q_{s,min}^2}{\Lambda_{QCD}^2}\right)\times \left[ 1+\frac{1}{2}ln\left(\frac{Q_{s,max}^2}{Q_{s,min}^2}\right)\left(1-\frac{Q_{s,max}}{\sqrt{s}}e^{|y|}\right)^4\right], 
\end{equation}
where $const$ is energy-independent, $Q_{s}^2 \equiv Q_{s}^2(s;y=0)$ and $Q_{s,min(max)}$ are defined as the smaller (larger) value of Eq. (43); at $y=0$, $Q_{s,min}^2=Q_{s,max}^2=Q_{s}^2(s)=Q_{s}^2(s_o)(s/s_o)^{\lambda/2}$.

Since $S_{A}Q_{s}^2 \sim N_{part}$, we can rewrite the above Eq.(44) as follows~\cite{[kharz]}
\begin{equation}
\frac{dN}{dy} = c~N_{part}\left(\frac{s}{s_o}\right)^{\lambda/2}e^{-\lambda (y)}\left[ln\left(\frac{Q_{s}^2}{\Lambda_{QCD}^2}\right)-\lambda |y|\right]\times \left[1+\lambda |y| \left(1-\frac{Q_s}{\sqrt{s}}e^{(1+\lambda/2)|y|}\right)^4\right], 
\end{equation}
with $Q_{s}^2(s)=Q_{s}^2(s_o)(s/s_o)^{\lambda/2}$.

\subsection{\bf{Model based on the percolation of strings}}
The physical situation described by the Glasma, the system of purely longitudinal fields in the region between the parting hadrons, is analogous to that underlying in the string percolation model (SPM). In the SPM one considers Schwinger strings, which can fuse and percolate~\cite{[string1],[string2],[string3],[string4]}, as the fundamental degrees of freedom. The effective number of strings, including percolation effects, is directly related to the produced particle's rapidity density. The SPM~\cite{[deus]} for the distribution of rapidity extended objects created in heavy-ion collisions combines the generation of lower center-of-mass  rapidity objects from higher rapidity ones with asymptotic saturation in the form of the well-known logistic equation for population dynamics:
\begin{equation}
\frac{\partial{\rho}}{\partial(-\Delta)}= \frac{1}{\delta}(\rho - A \rho^2),
\end{equation} 
where $\rho$ $\equiv$ $\rho(\Delta,Y)$ is the particle density, Y is the beam rapidity, and $\Delta \equiv \left|y\right| - Y$. The variable -$\Delta$ plays the role of evolution time, parameter $\delta$ controls the low density $\rho$ evolution and parameter A responsible for saturation. The Y-dependent limiting value of $\rho$, determined by the saturation condition $\frac{\partial\rho}{\partial(-\Delta)}$=0, is given by $\rho_{Y}$=$\frac{1}{A}$. Whereas the separation between the region $\Delta > \Delta_0$ (i.e, low density and positive curvature) and the region $\Delta < \Delta_0$ (i.e, high density and negative curvature) is defined by $\frac{\partial^2 \rho}{\partial (-\Delta)^2}\vert_{\Delta_0}$=0, which  gives $\rho_0 \equiv \rho(\Delta_0,Y) = \frac{\rho Y}{2}$. Integrating Eq. (46) between $\rho_0$ and some $\rho(\Delta)$, one can obtain~\cite{[deus]}:
\begin{equation}
\rho(\Delta,Y) = \frac{\rho Y}{e^{\frac{\Delta - \Delta_0}{\delta}}+1}.
\end{equation} 
The particle density in SPM is proportional to $N_A$ (i.e, $\rho \propto N_A$), therefore we can write
\begin{equation}
\rho(\Delta,Y, N_A) = \frac{N_A\rho Y}{e^{\frac{\Delta - \Delta_0}{\delta}}+1}.
\end{equation}  
To be more specific regarding the quantities $\delta, \rho_Y$ and $\Delta_0$, $\delta$ does not strongly depend on $Y$, the parameter $\rho_Y$ is the normalized particle density at mid-rapidity and is related to gluon distribution at small $x$ i.e, $\rho \propto e^{\lambda Y}$. As $\rho_Y$ increases with rapidity Y, energy conservation arguments gives $\Delta_0 = -\alpha Y$ with $0 < \alpha <1$. Now Eq. (48) can be rewritten as
\begin{equation}
\frac{1}{N_A}\rho(\Delta,Y) \equiv \frac{2}{N_{part}}\frac{dn}{dy}= \frac{e^{\lambda Y}}{e^{\frac{\Delta + \alpha Y}{\delta}}+1}.
\end{equation}
The particle density $\rho$ is a function of both $\Delta$ and the beam rapidity $Y$ whereas according to the hypothesis of limiting fragmentation, for $\Delta$ larger than some $Y$-dependent threshold, the density $\rho$ remains a function of $\Delta$ only. 
In the Glauber model, energy-momentum conservation constrains the combinatorial factors at low energy but in the framework of SPM energy-momentum conservation is accounted by reducing the effective number of sea strings via
\begin{equation}
N_{A}^{4/3} \to N_{A}^{1+\alpha(\sqrt{s})},
\end{equation} 
where $N_A = N_{part}/2$  with 
\begin{equation}
\alpha(\sqrt{s})=\frac{1}{3}\left( 1-\frac{1}{1+ln(\sqrt{s/s_0} + 1)} \right).
\end{equation} 
which yields
\begin{equation}
{\frac{1}{N_{A}}\frac{dn_{ch}^{AA}}{dy}}{\vert}_{y=0} \sim N_{A}(N_{A}^{\alpha(\sqrt{s})}-1)\frac{dn^{pp}}{dy}{\vert}_{y=0}
\end{equation} 
In the string percolation model (SPM), the charged particle multiplicity at mid-rapidity in proton-proton interactions is given by~\cite{[bautista]}:
\begin{equation}
{\frac{dn_{ch}^{pp}}{d\eta}}{\vert}_{\eta=0} = kF(\eta^{t}_{p})N^{s}_{p}
\end{equation} 
such that for symmetric nucleus-nucleus case, one has~\cite{[bautista]}:
\begin{equation}
{\frac{1}{N_{A}}\frac{dn_{ch}^{AA}}{d\eta}}{\vert}_{\eta=0} = \kappa F(\eta^{t}_{p})N^{s}_{p}\left(1 + \frac{F(\eta^{t}_{N_A})}{F(\eta^{t}_{p})}(N_{A}^{\alpha(\sqrt{s})}-1)\right),
\end{equation} 
Above Eq. (54) involves three parameters, the normalization constant $k$; the threshold scale $\sqrt{s_0}$ in $\alpha(\sqrt{s})$; the power $\lambda$. The values of fitting parameters obtained from the best fit to pp and AA data are: $k$ = 0.63$\pm$0.01, $\lambda$=0.201$\pm$0.003 and the threshold scale $\sqrt{s_0}$=245$\pm$29 GeV. In the SPM, the power law dependence of the multiplicity on the collision energy is the same in pp and AA collisions. 

The pseudo-rapidity dependence of AA collisions is given by~\cite{[ira.bautista]}:
\begin{equation}
{\frac{1}{N_{A}}\frac{dn_{ch}^{AA}}{d\eta}}{\vert}_{\eta} =  \kappa F(\eta^{t}_{p})N^{s}_{p}\frac{\left(1 + \frac{F(\eta^{t}_{N_A})}{F(\eta^{t}_{p})}(N_{A}^{\alpha(\sqrt{s})}-1)\right)}{exp\left(\frac{\eta-(1-\alpha)Y}{\delta}+1\right)},
\end{equation} 
where $J$ is the Jacobian $J=\frac{cosh\eta}{k_1 + sinh^{2}\eta}$ and ${\kappa}'= \frac{\kappa}{J(\eta=0)}(exp(\frac{-(1-\alpha)Y}{\delta})+1) $.
\begin{figure}
\begin{center}
\includegraphics[width=15 cm]{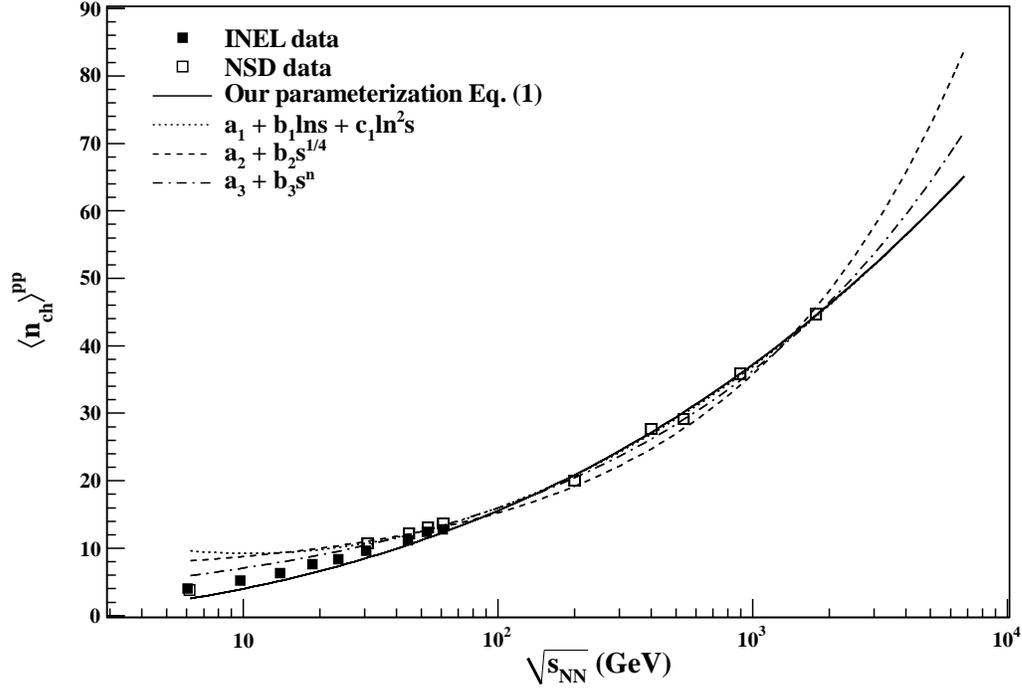}
\end{center}
\caption{Variations of total mean multiplicities of charged hadrons ($N_{ch}$) produced in $p-p$ collisions as a function of c.m energy ($\sqrt{s_{NN}}$).  Figure is taken from Ref.~\cite{[ashwini]}. }
\label{fig:1} 
\end{figure}

\begin{figure}
\begin{center}
\includegraphics[width=15 cm]{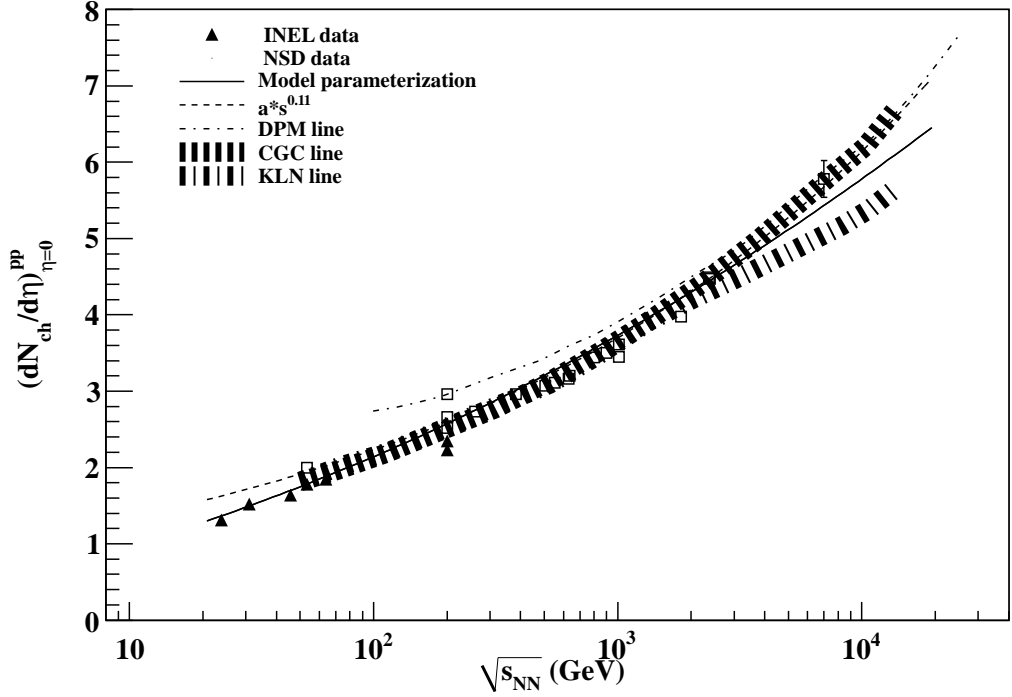}
\end{center}
\caption{ Variations of the charged hadrons pseud-rapidity density at mid-rapidity in $p-p$ collisions as a function of c. m. energy. Data is taken from Ref.~\cite{[thome],[arnison],[albajar],[ansorge],[alner1],[alner2],[abe],[sun],[noucier],[abelev],[aamodt],[aamodt1],[khachatryan1],[khachatryan2]}. }
\label{fig:1} 
\end{figure}

\begin{figure}
\begin{center}
\includegraphics[width=15 cm]{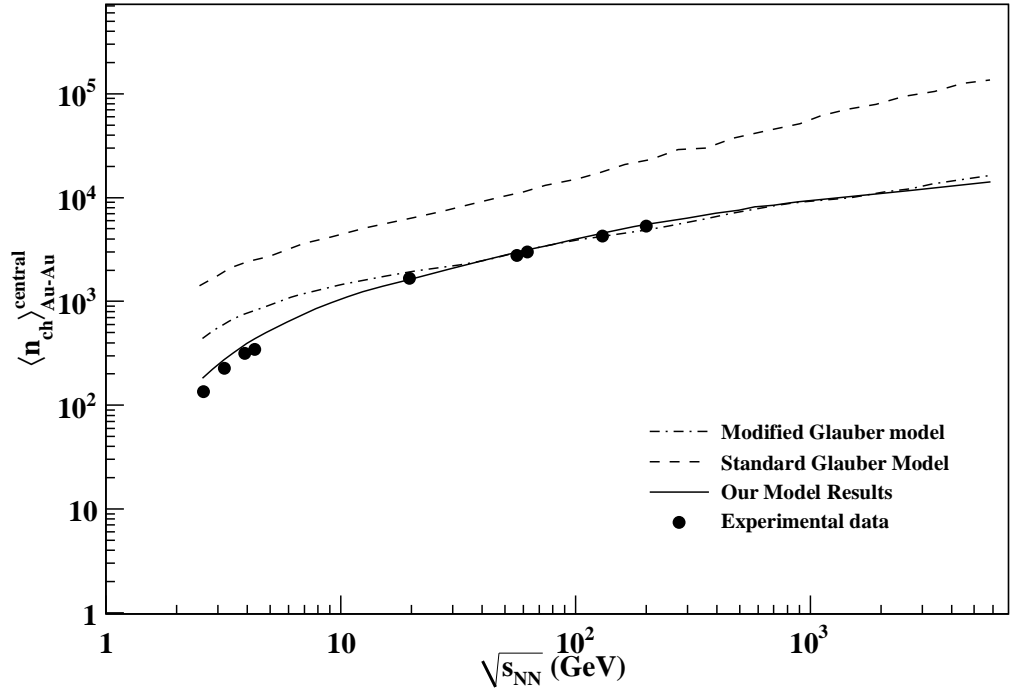}
\end{center}
\caption{Variations of total mean multiplicity of charged hadrons in central $Au-Au$ collisions with $\sqrt{s_{NN}}$. Figure is taken from Refs.~\cite{[ashwini]}.   }
\label{fig:1} 
\end{figure}

\begin{figure}
\begin{center}
\includegraphics[width=15 cm]{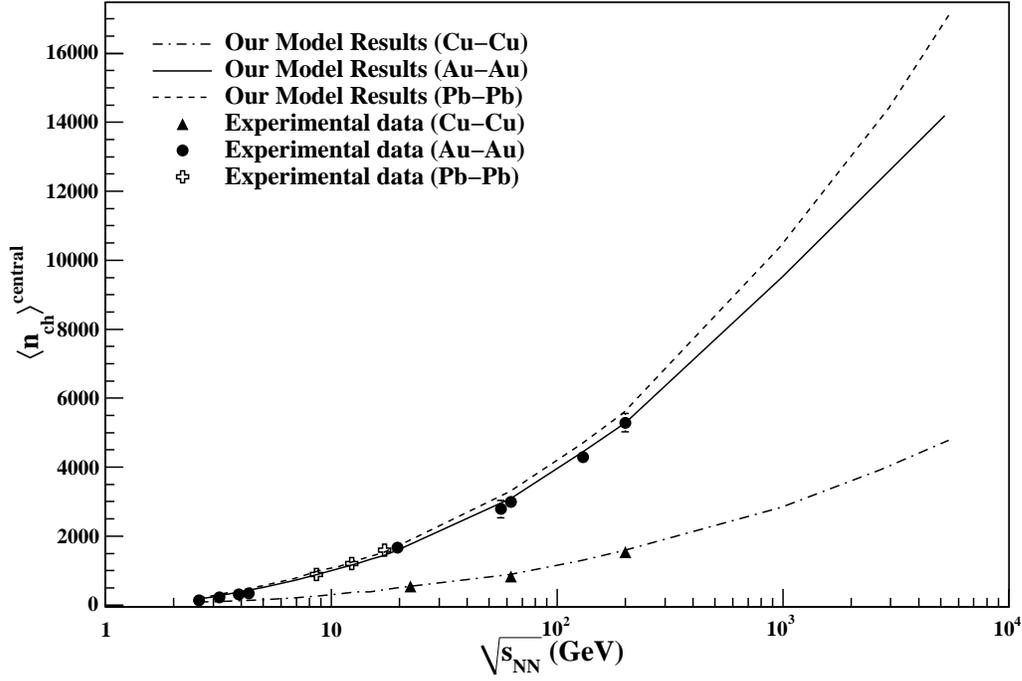}
\end{center}
\caption{Total number of charged hadrons ($N_{ch}$) produced in $Au-Au, Cu-Cu$ and $Pb-Pb$ collisions as a function of c. m. energy. Figure is taken from Ref.~\cite{[ashwini]}.}
\label{fig:1} 
\end{figure}

\begin{figure}
\begin{center}
\end{center}
\label{fig:4} 
\end{figure}

\section{Charged Hadron Multiplicity Distributions }
Here, we discuss various observed features of the measured charged hadron multiplicity in hadron-hadron, hadron-nucleus and nucleus-nucleus interactions at different energies and their comparison in the view of above mentioned models.

\subsection{\bf{Mean Multiplicity as a function of $\sqrt{s_{NN}}$}}
Feynman was first to point out that the multiplicity spectrum observed in proton-proton collisions at asymptotically large energies becomes independent of the center-of-mass energy($\sqrt{s}$) as $\sqrt{s} \to \infty$~\cite{[feynman],[benecke],[hagedron]}. This assumption is known as $Feynman~ scaling$. He concluded this fact primarily based on the phenomenological arguments about the exchange of quantum numbers between the colliding particles. This energy independent behaviour of the height of rapidity distribution around mid-rapidity naturally implies that the total multiplicity after integration over rapidity involves $ln\sqrt{s}$ behaviour since $y_{max} = ln \frac{\sqrt{s}}{m_N}$, where $m_N$ is the nucleon-mass. However up to $\sqrt{s} = 1800$ GeV the experimental data does not indicate that the height of rapidity distribution around mid-rapidity (i.e, $(\frac{dn}{d\eta})_{\eta=0}$) gets saturated. Thus, Feynman scaling is violated by continuous increase in the central rapidity density. Later on, it was realized that the gluons arising from gluon-bremsstrahlung  processes give QCD-radiative corrections~\cite{[gunion]}. Recently, it was noticed by PHOBOS Collaboration from the $p-p$ and/or $p-\bar{p}$ data that central plateau height i.e., $(\frac{dn}{d\eta})_{\eta =0}$ grows as $ln^{2}{\sqrt{s}}$ which in turn will give $ln^{3}{\sqrt{s}}$ type behaviour in the total multiplicity~\cite{[back1],[jeon],[back2]}. Thus, the violation of Feynman scaling  clearly indicates that this scaling type of behaviour is not supported by experiments.  Based on the above findings, we have included $ln^{3}{\sqrt{s}}$ behaviour as in Eq. (15) and have shown  its applicability in describing the $p-p$ and $p-\bar{p}$ data quite successfully for the available entire energy range till date.

In Fig. 1, the inelastic (filled symbols) and non-single diffractive (NSD) data (open symbols) of charged hadron multiplicity in full phase space for $p-p$ collisions at various center-of-mass energies from different experiments e.g., ISR, UA5 and E735 is shown. Inelastic data at very low energies (filled symbols) are used because NSD data are not available for these energies and also the trend shows that the difference between inelastic and NSD data is very small at lower energies. Further,  this data set is fitted with  three different functional forms in order to make a simultaneous comparison with our parameterization as given in Eq. (15). The short-dashed line has the functional form as : $a_2+b_2s^{1/4}$ which is actually inspired by the Fermi-Landau model. It provides a resonable fit to the data at higher $\sqrt{s_{NN}}$ with $a_2 = 5.774$ and $b_2 = 0.948$~\cite{[grosse]}. However, since $a_2$ summarizes the leading particle effect also, its value  should not be much larger than two. The dotted line represents the functional form as : $a_1+b_1$ $ln s+c_1$ $ln^{2}s$ and it fits the data well at higher $\sqrt{s_{NN}}$ but shows a large disagreement with the experimental data at lower center-of-mass energies with the values  $a_1 = 16.65$, $b_1 = -3.147$ and $c_1 = 0.334$, respectively. The dashed-dotted line represents the form $a_3+b_3s^{n}$ and it provides a qualitative description of the data with $a_3=0, ~b_3=3.102$ and $n=0.178$ ~\cite{[grosse]}. The solid line represents the parametrization given by Eq. (15) according to Ref.~\cite{[ashwini]}. In Fig. 2, we present the inelastic (filled symbols) and non-single diffractive (NSD) data (open symbols) of $\frac{dn_{ch}}{d\eta}$ at mid-rapidity for $p-p$ collisions at various center-of-mass energies from different experiments e.g., ISR, UA5, E735, RHIC and LHC~\cite{[thome],[arnison],[albajar],[ansorge],[alner1],[alner2],[abe],[sun],[noucier],[abelev],[aamodt],[aamodt1],[khachatryan1],[khachatryan2]}. Solid line represents the parameterization used in Eq.(19) according to Ref.~\cite{[ashwini]}. DPM results~\cite{[dpm1]} for charged hadron pseudo-rapidity densities are shown by the dashed-dotted line for $p-p$ collisions. In DPM, at lower energiers a larger value of multiplicities is obtained which is not supported by the experimental data which shows a $s^{0.11}$ type of energy dependent behaviour (shown by dashed line) for the entire range of energy including LHC energy fits the data well. The results on hadron production in $p-p$ collisions by using a saturation approach based on the CGC formalism are shown by the dotted line with error bands while KLN model results are shown by dashed-dotted line with same error bands~\cite{[rezaeian]}. Both the approaches are based on $k_T$ factorization, while the main difference lies in determining the saturation scale for the nucleus~\cite{[rezaeian]}. 

Fig. 3 shows the variation of mean multiplicity of charged hadrons produced in central $Au-Au$ collision with respect to $\sqrt{s_{NN}}$. We also compare our model results (solid line) with the standard Glauber model (dashed line) and the modified Glauber model results (dash-dotted line) ~\cite{[feofilov]} alongwith the experimental data of AGS and RHIC~\cite{[adler],[backau-au],[klay]}. We would like to mention here the main difference between modified Glauber model and the standard Glauber calculation. In standard Glauber model, a participating nucleon can have one or more number of collision at the same $\sqrt{s_{NN}}$ and  a constant value for $\sigma_{inel}^{pp} (\sqrt{s})$ is used in the calculation. However, in modified Glauber model, a participating nucleon loses a fraction of its momentum after the first collision and participate in subsequent collisions with somewhat lower energy and also a lower value of $\sigma_{inel}^{pp} (\sqrt s)$ is used in the calculation. This modification suppresses the number of collisions significantly in comparison to that obtained in the standard Glauber model. We find that the results obtained in our model give a better description to the experimental data in comparison to the standard as well as modified Glauber model predictions~\cite{[ashwini]}. This clearly shows the significance of the role played by quark degrees of freedom as used by us in our picture~\cite{[ashwini]}.
In Fig. 4, the variations of mean multiplicity of charged hadrons produced in $Au-Au, Cu-Cu$ and $Pb-Pb$ collisions as a function of $\sqrt{s_{NN}}$ in the wounded quark model is shown alongwith the comparison of the experimental data~\cite{[alver],[adler],[backau-au],[klay]}.

\subsection{\bf{Charged hadron multiplicity as a function of centrality}}

\begin{figure}
\begin{center}
\includegraphics[width=15 cm]{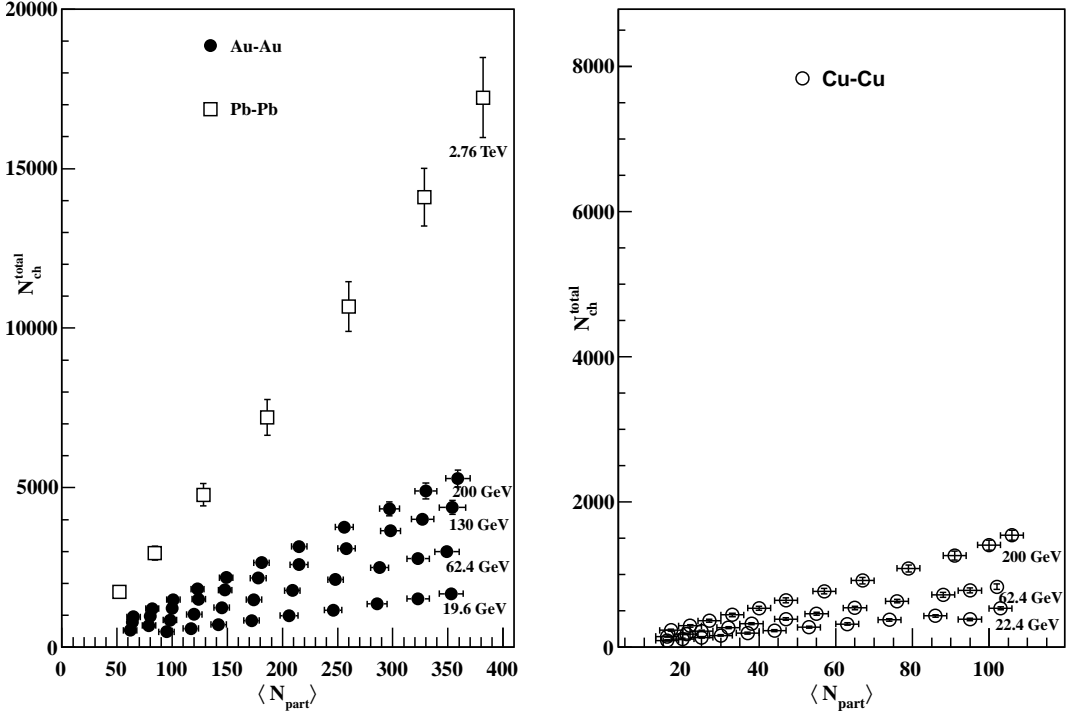}
\end{center}
\caption{Total number of charged hadrons ($N_{ch}$) as a function of centrality ($\langle N_{part}\rangle$). The experimental data has been taken from Ref.~\cite{[alver],[toia]}.  }
\label{fig:1} 
\end{figure}

\begin{figure}
\begin{center}
\includegraphics[width=15 cm]{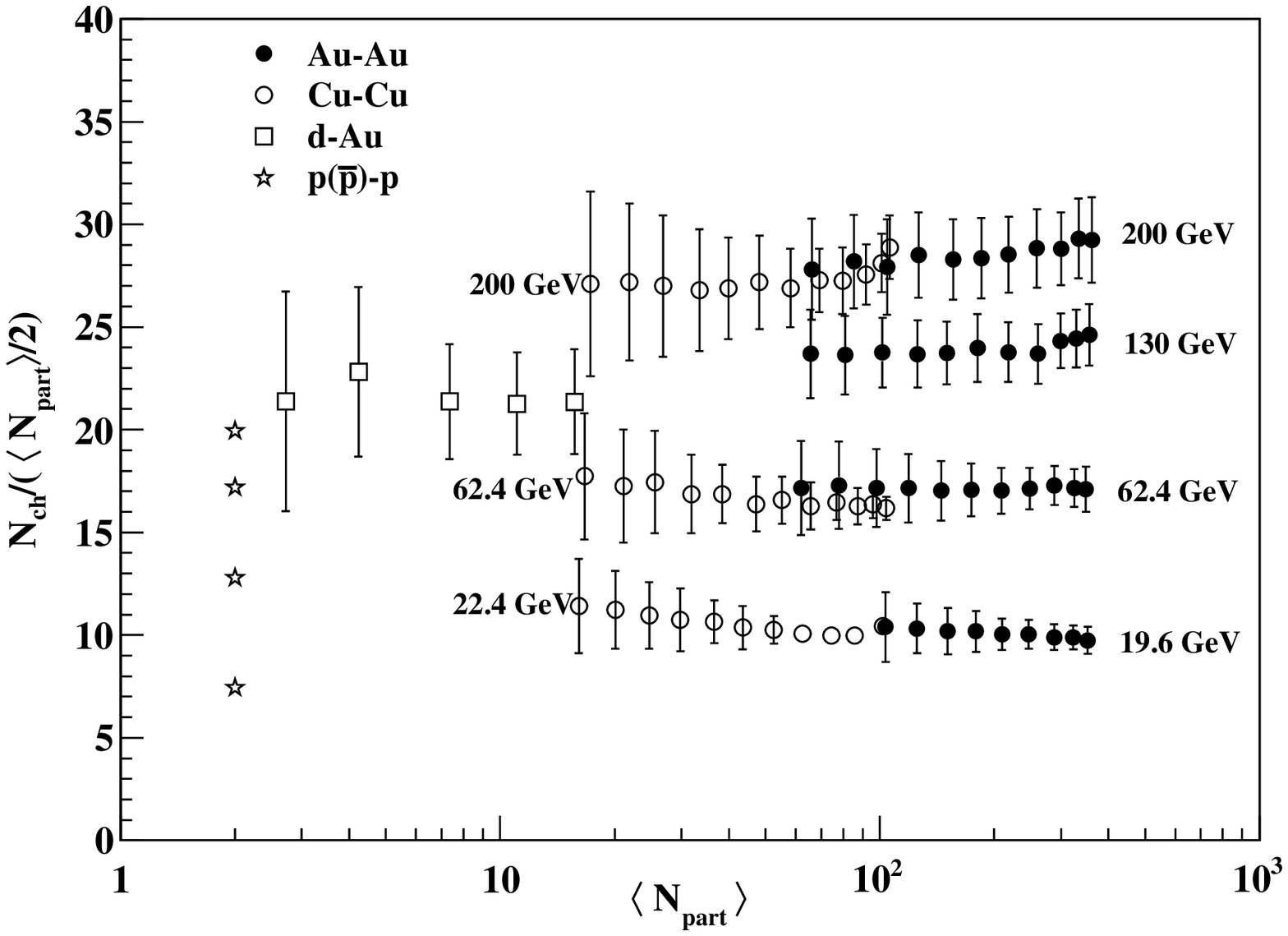}
\end{center}
\caption{ Charged hadron multiplicity per participant pair ($N_{ch}/{\langle N_{part}/2 \rangle}$)  as a function of $N_{part}$ at different $\sqrt{s_{NN}}$ for A-B collisions~\cite{[b.b.backet.al.],[backau-au],[backd-au],[backcu-cu]}. Inelastic $p(\bar{p})-p$ data or interpolations to unmeasured energies with $N_{part}=2$ taken from Ref.~\cite{[back],[back11],[b.b.backet.al.]} }
\label{fig:1} 
\end{figure}

The total charged hadron multiplicity in hadronic collisions is of great significance as it provides a direct measure of the degrees of freedom released in the collision process.   The role of the system-size dependence on the particle production in high energy nuclear collisions is of prime interest and is mainly studied either by varying the size of the colliding nuclei or by varying the centrality of the collision event. This variation of the collision centrality directly affects the volume of the collision zone formed in the initial stage and it also affects the number of binary collisions suffered by each nucleon. Furthermore, the study of centrality dependence along with the varying collision energy also reflects the contribution of the soft and hard process involved in the particle production mechanism. In Fig. 5, a compilation of the measured total charged hadron multiplicity as a function of centrality ($\langle N_{part} \rangle$) in nucleus-nucleus ($A-A$) collisions at varying energies is shown. Experimental data clearly shows a non-trivial growth of multiplicity with atomic number of colliding nuclei and also the multiplicity dependence on the collision centrality and energy is clearly visible. The main feature of total particle production in $Au-Au, Cu-Cu$ and $Pb-Pb$ collisions is a direct proportionality of the total charged hadron multiplicity to the number of pair of participant nucleons $N_{part}$. This proportionality becomes stronger with increase in collision energy (as observed in the form of increasing slopes of the curve) for each colliding system.   Total multiplicity of charged hadron per participant pair as a function of centrality  in $d-Au, Au-Au ~\&~ Cu-Cu$ collisions at RHIC energies are shown  in Fig. 6. The total charged hadron multiplicity scales with $\langle N_{part}\rangle$ in $d-Au, Au-Au ~\&~ Cu-Cu$ collisions indicating that the transition between inelastic $p(\bar{p})-p$ and $A-A$ collisions is not controlled simply by the number of participants, as even very central $d-Au$ collisions do not show any sign of trending up towards the level of the $Au-Au$ data~\cite{[b.back]}.

\begin{table*}
\begin{center}
\caption{The total charged hadron multiplicities as a function of centrality in Au-Au Collisions at three different RHIC energies~\cite{[ashwini],[b.b.backet.al.]}. }
\begin{tabular}{|l|l|l|l|l|l|l|l|l|l|l|l|l}
\hline
{Centrality Bin} & \multicolumn{2}{l|}{{$\sqrt{s_{NN}}=200$ GeV}}&\multicolumn{2}{l|}{{$\sqrt{s_{NN}}=130$ GeV}} &\multicolumn{2}{l|}{{$\sqrt{s_{NN}}=62.4$ GeV}}  \\
\cline{2-7}
 &{Model}&{Experimental }&{Model }&{Experimental }&{Model }&{Experimental }\\
\hline\hline

  ~$0-6 \%$  & 5277     &5095$\pm$255   & 4449       & 4195$\pm$210     &3098     &2881$\pm$143           \\
 ~$6-10 \%$  & 4177     &4341$\pm$245   & 3480       & 3649$\pm$182       &2354     &2489 $\pm$124             \\
$10-15 \%$   & 3601     &3763$\pm$188   & 3072       & 3090$\pm$155       &2196     &2120$\pm$106           \\
$15-20 \%$   & 2939     &3153$\pm$158   & 2554       & 2586$\pm$129       &1886     &1777$\pm$88                \\
$20-25 \%$   & 2492     &2645$\pm$132   & 2189       & 2164$\pm$108       &1673     &1485$\pm$74             \\
$25-30 \%$   & 2003     &2184$\pm$109   & 1779       & 1793$\pm$90        &1392     &1236$\pm$61              \\
$30-35 \%$   & 1790     &1819$\pm$91    & 1596       & 1502$\pm$75        &1262     &1027$\pm$51              \\
$35-40 \%$   & 1591     &1486$\pm$74    &1428        & 1222$\pm$61        &1104     &840$\pm$42            \\
$40-45 \%$   & 1230     &1204$\pm$60    & 1111       & 975$\pm$49         &902      &679$\pm$33               \\ 
$45-50 \%$   & 881      &951$\pm$48     & 801        & 782$\pm$39        &662      &532$\pm$26 \\ \hline
\end{tabular}
\end{center}
\end{table*}

\begin{table*}
\begin{center}
\caption{The total charged hadron multiplicities as a function of centrality in Cu-Cu Collisions at three different RHIC energies~\cite{[ashwini],[b.b.backet.al.]}.}
\begin{tabular}{|l|l|l|l|l|l|l|l|l|l|l|l|l}
\hline
{Centrality Bin} & \multicolumn{2}{l|}{{$\sqrt{s_{NN}}=200$ GeV}}&\multicolumn{2}{l|}{{$\sqrt{s_{NN}}=62.4$ GeV}} &\multicolumn{2}{l|}{{$\sqrt{s_{NN}}=22.4$ GeV}}  \\
\cline{2-7}
 &{Model}&{Experimental }&{Model }&{Experimental }&{Model }&{Experimental }\\
\hline\hline

  ~$0-6 \%$  & 1532       &1474$\pm$69  & 884         &807$\pm$35       &483     &508$\pm$22    \\
 ~$6-10 \%$  & 1203       &1262$\pm$59  & 654         &721$\pm$32       &395     &431$\pm$19             \\
$10-15 \%$   & 1003       &1084$\pm$51  & 609         &635$\pm$27       &355     &375$\pm$18         \\
$15-20 \%$   & 850        &917$\pm$43   & 544         &541$\pm$24       &339     &320$\pm$15         \\
$20-25 \%$   & 708        &771$\pm$38   & 470         &460$\pm$21       &308     &273$\pm$14           \\
$25-30 \%$   & 600        &645$\pm$32   & 411         &386$\pm$17       &279     &230$\pm$13            \\
$30-35 \%$   & 504        &538$\pm$27   & 354         &323$\pm$15       &247     &194$\pm$12              \\
$35-40 \%$   & 400        &444$\pm$23   & 288         &270$\pm$13       &207     &162$\pm$12            \\
$40-45 \%$   & 325        &364$\pm$19   & 239         &223$\pm$11       &175     &135$\pm$11               \\ 
$45-50 \%$   & 270        &293$\pm$15   & 201         &183$\pm$9        &150     &112$\pm$11 \\ \hline

\end{tabular}
\end{center}
\end{table*}
In Table 1 and Table 2, the wounded quark model~\cite{[ashwini]} results are organized in tabular form for ten centrality classes and are compared accordingly with the RHIC multiplicity data~\cite{[b.b.backet.al.]}. For suitable comparison of model results for most central ($0-6\%$) case in Au-Au and Cu-Cu collisions, we have taken the average of the RHIC multiplicitiy data~\cite{[b.b.backet.al.]} of first two centrality bins ($0-3\% ~ \& ~ 3-6\%$).  In Table 1,  a tabular representation of wounded quark model results for the total charged hadron multiplicities as a function of the centrality in Au-Au collisions at three RHIC energies ($\sqrt{s_{NN}}$ = 62.4, 130 and 200 GeV). We have also made a comparison with the experimental data~\cite{[b.b.backet.al.]} at the respective RHIC energies and find a reasonable agreement within experimental errors. Similarly in Table 2,  wounded quark model results for the total charged hadron multiplicities as a function of the centrality in Cu-Cu collisions at three RHIC energies ($\sqrt{s_{NN}}$ = 22.4, 62.4 and 200 GeV) are shown and compared with the experimental data~\cite{[b.b.backet.al.]} at the respective RHIC energies. Thus, the centrality dependence of charged hadron production involving different colliding nuclei ( Au-Au and Cu-Cu ) at RHIC energies is well described by wounded quark model which clearly suggests that it can be used reliably in explaining the multiplicity data for other colliding nuclei too.   
\subsection{\bf{Charged hadron pseudo-rapidity distributions }}
The pseudo-rapidity distribution of charged hadrons is another quantity in the studies of the particle production mechanism in high energy hadron-hadron and nucleus-nucleus collisions. Pseudo-rapidity density is a well defined quantity which is sensitive to the initial conditions of the system i.e., parton shadowing and the effects of rescattering and hadronic final state interactions. Fig.7 shows the pseudo-rapidity distributions of charged hadrons produced in the most central $Au-Au$ collisions over the entire range of $\eta$ at $\sqrt{s_{NN}}$ = 19.6, 62.4, 130 and 200 GeV and in Pb-Pb collisions at 2.76 TeV. CGC model results at 130 GeV, 200 GeV (for Au-Au collisions) and 2.76 TeV (for Pb-Pb collisions)~\cite{[amir]} are shown with dotted lines. In this approach, the charged hadron pseudo-rapidity distributions in nucleus-nucleus collisions are calculated within $k_T$ factorization. CGC model results show a good agreement for Au-Au collisions at RHIC energies (130 GeV and 200 GeV) whereas for Pb-Pb collisions at LHC energy near mid-rapidity it successfully reproduces the data within limited pseudo-rapidity range ($|\eta| < 2$).   SPM results~\cite{[ira.bautista]} are shown with dashed line for $Cu-Cu,~Au-Au~\& Pb-Pb$ collisions at different energies in Fig. 7 and Fig.8.  In SPM approach, there are three different regions involved for pseudo-rapidity  distribution. In the fragmentation regions, there are only strings involving valence quarks while in the central region there are additional short rapidity strings between quarks and  antiquarks. The nonequilibrium statistical relativistic diffusion model (RDM) results~\cite{[wolschin]} at RHIC and LHC energies are shown in Fig. 7, which gives a reasonable discription of $AA$ data. In RDM there are also three sources, one at central rapidity and the two others in the fragmentation regions. Solid lines in Fig. 7 and Fig.8 are results obtained in the wounded quark model~\cite{[ashwini]} for $Au-Au, Cu-Cu$ collisions at RHIC energies  and for $Pb-Pb$ collisions at LHC energy. Wounded quark model quite successfully explains the data on the pseudo-rapidity distribution for all the three colliding systems (Au-Au, Cu-Cu and Pb-Pb) at RHIC and LHC energies.
\begin{figure}
\begin{center}
\includegraphics[width=15 cm]{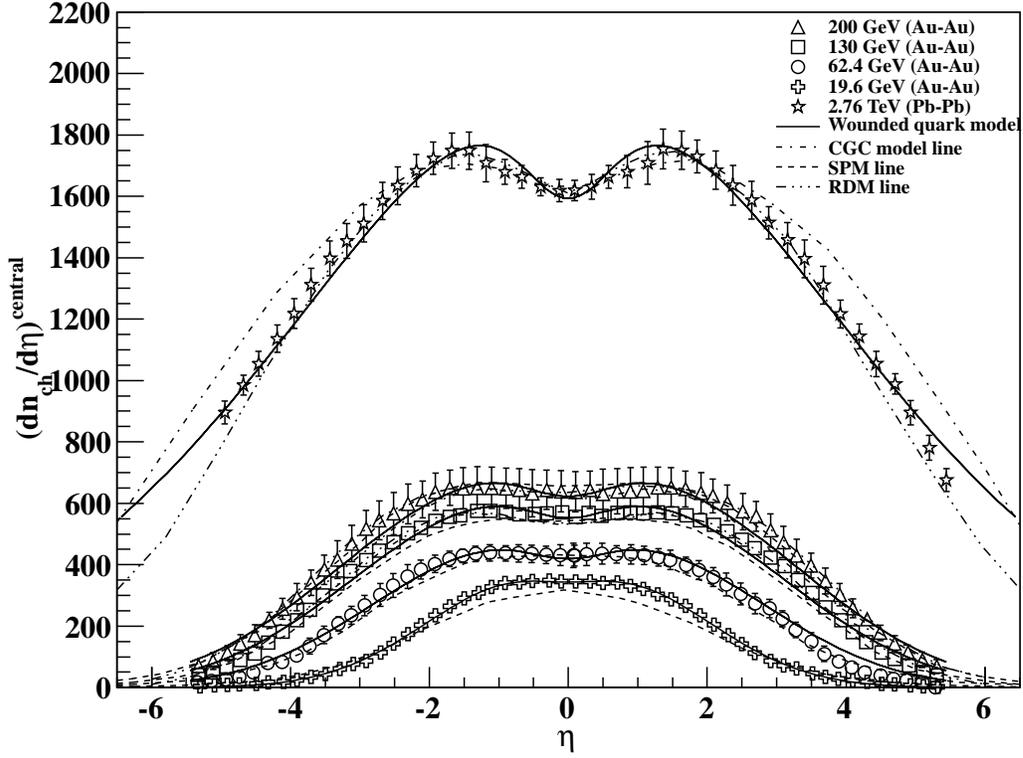}
\end{center}
\caption{Charged hadron pseudo-rapidity distribution in $Au-Au$ collisions at different RHIC energies and $Pb-Pb$ at LHC energy along with comparison of different model results. Data are taken from Ref.~\cite{[b.b.backet.al.],[gulbrandsen],[amir],[toia]}.   }
\label{fig:5} 
\end{figure}

\begin{figure}
\begin{center}
\includegraphics[width=15 cm]{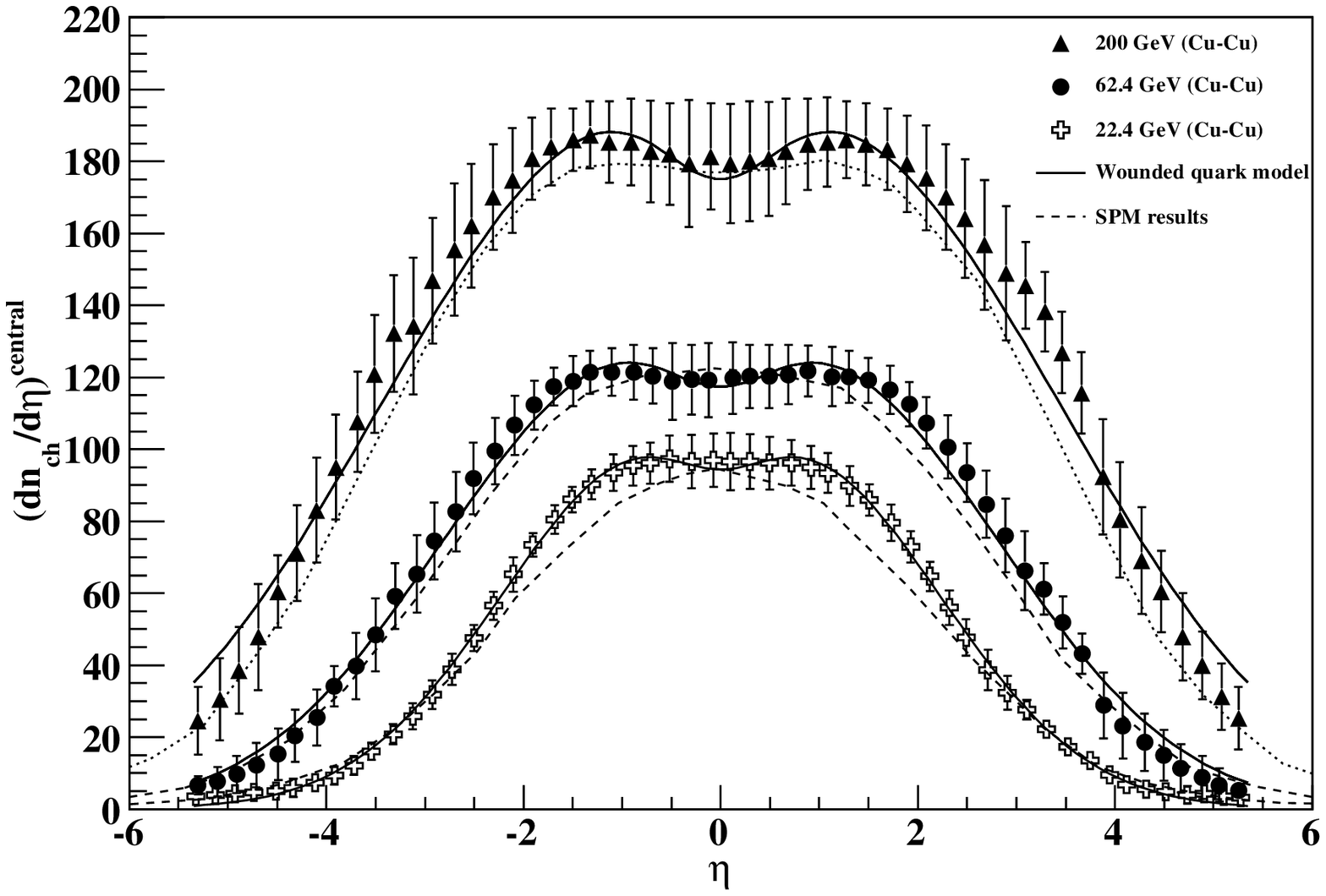}
\end{center}
\caption{Charged hadron pseudo-rapidity distribution in Cu-Cu collisions at different RHIC energies along with comparison of different  model results. Data are taken from Ref.~\cite{[veres]}.   }
\label{fig:5} 
\end{figure}

\begin{figure}
\begin{center}
\includegraphics[width=15 cm]{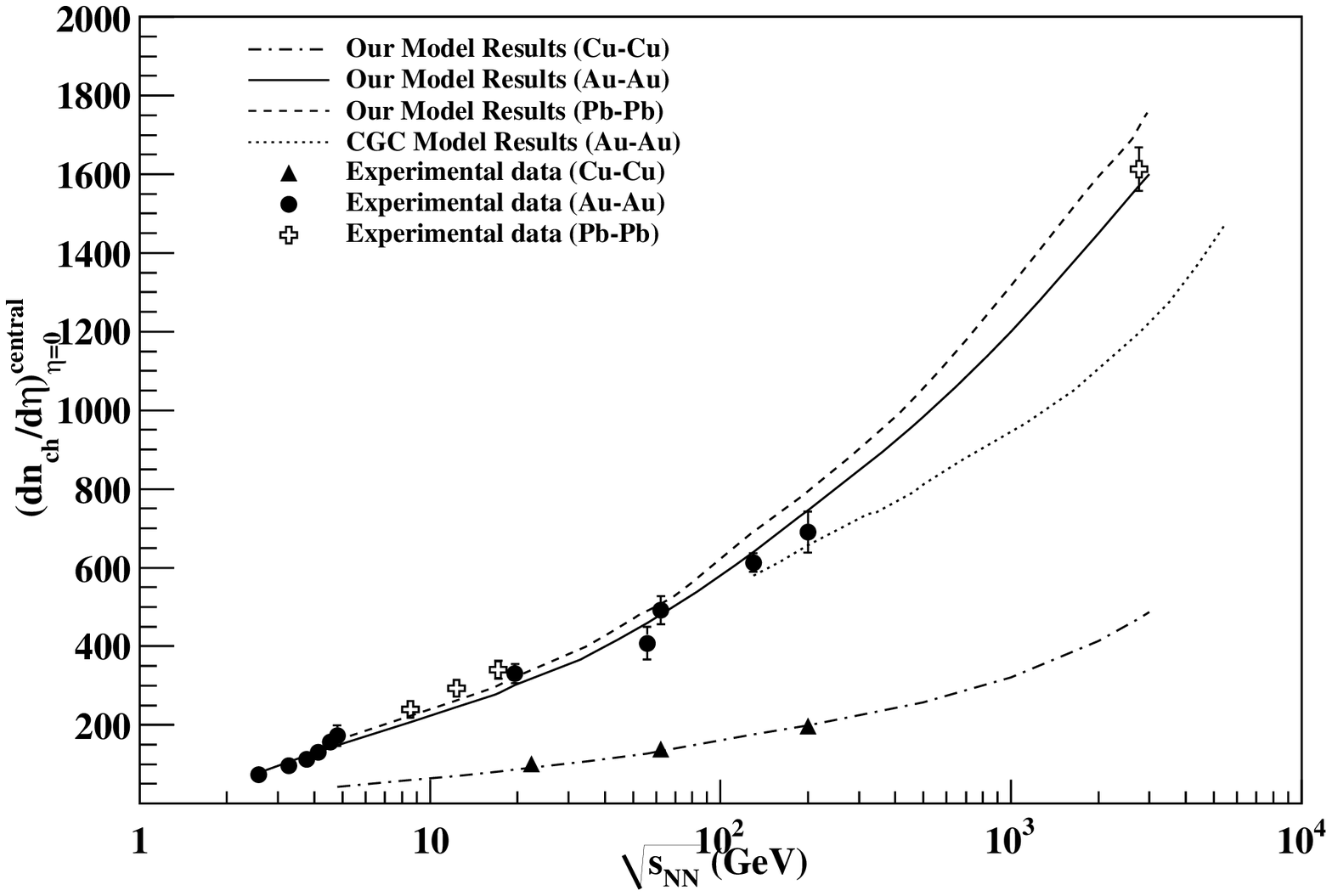}
\end{center}
\caption{Variations of pseudo-rapidity density of charged hadrons at mid-rapidity as a function of $\sqrt{s_{NN}}$ for different colliding nuclei along with the comparison of different model results~\cite{[ashwini],[j.l.albacete]}.} 
\label{fig:2} 
\end{figure}
\subsection{\bf{Central Pseudo-rapidity density as a function of $\sqrt{s_{NN}}$ as well as colliding nuclei}}

 Pseudo-rapidity density of charged hadrons provides relevant information on the temperature ($T$) as well as the energy density of the QGP. Further more, the study of the dependence of charged hadron densities at mid-rapidity with the c. m. energy and system size can provide the relevant information on the interplay between hard parton-parton scattering process and soft proesses. 
 At first sight, it looks logistic to consider the nucleus-nucleus collision as an incoherent superposition of nucleon-nucleon collisions as in the framework of wounded nucleon model approach. However, recent results on multiplicity data at RHIC and LHC energies for proton-proton and nucleus-nucleus collisions give an indication that nucleus-nucleus collisions are not simply an incoherent superposition of the collisions of the participating nucleons as $(\frac{dn_{ch}^{AA}}{d\eta})_{\eta=0}> N_{part}.(\frac{dn_{ch}^{pp}}{d\eta})_{\eta=0}$. It hints towards the role of multiple scattering in the nucleon-nucleon collisions. Further, the scaling with number of binary collisions also does not seem to hold good as $(\frac{dn_{ch}^{AA}}{d\eta})_{\eta=0}<< N_{coll}.(\frac{dn_{ch}^{pp}}{d\eta})_{\eta=0}$, which indicates towards the coherence effects involved in these collision processes~\cite{[ir.bautista]}. At higher energies, the role of multiple scattering~\cite{[b.b.b.back]} becomes more significant in describing the nucleus-nucleus collisions due to the contribution of hard processes. In Fig.9, the pseudo-rapidity densities of charged hadrons at mid-rapidity is shown with c. m. energy ($\sqrt{s_{NN}}$) for different colliding nuclei.  Wounded quark model results (as obtained from Eq. (20)) which is based on the two component model are shown along with the experimental data and are in quite good agreement. It clearly signifies the role of hard and soft process involved in these collision processes. CGC model results for Au-Au collisions is shown with dotted line~\cite{[j.l.albacete]}.

\begin{figure}
\begin{center}
\includegraphics[width=15 cm]{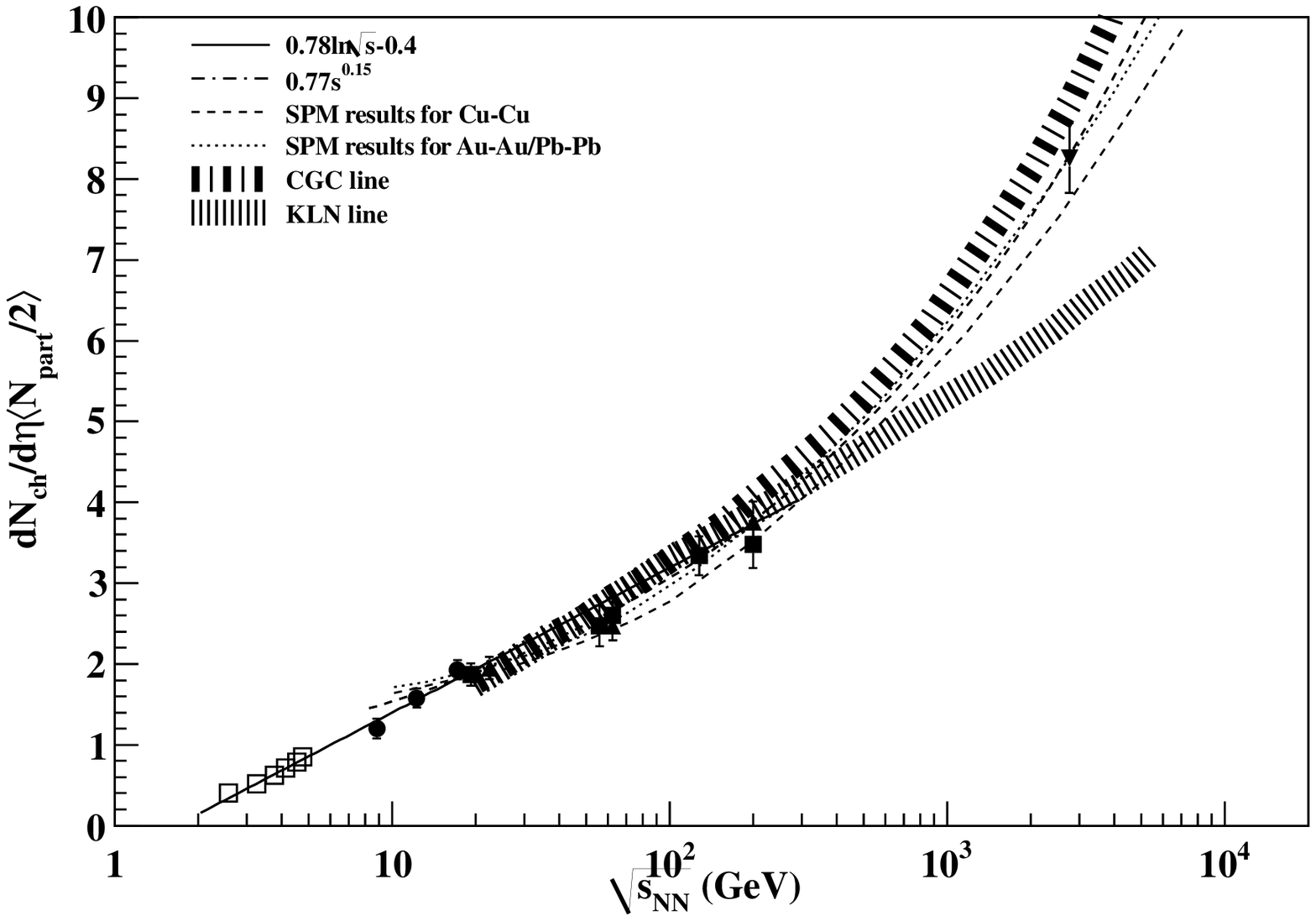}
\end{center}
\caption{Energy dependence of  of charged hadron mid-pseudorapidity density per participant pair. SPM~\cite{[bautista]}, CGC and KLN model~\cite{[rezaeian]} results are shown. }
\label{fig:3} 
\end{figure}
\begin{figure}
\begin{center}
\includegraphics[width=15 cm]{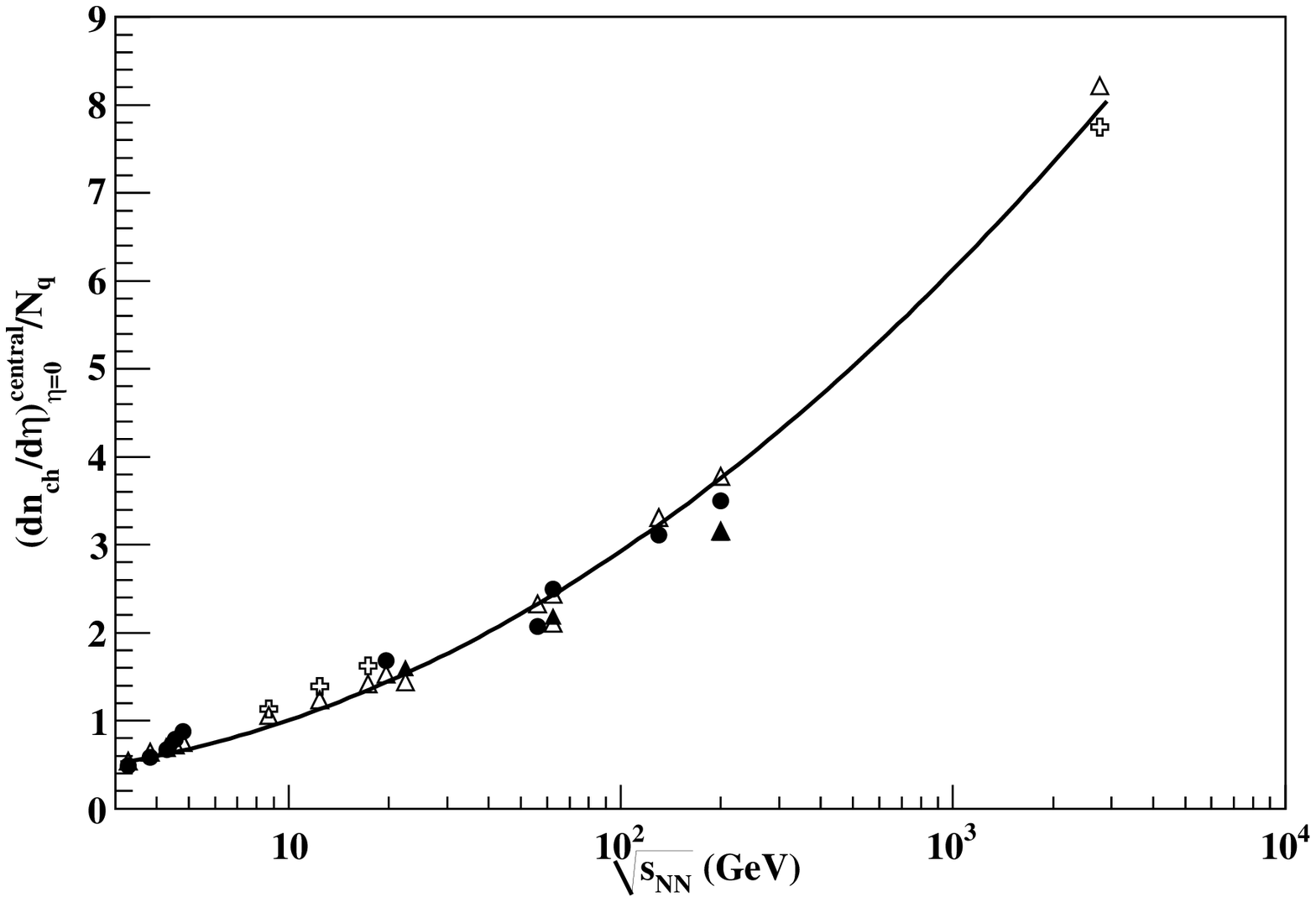}
\end{center}
\caption{Energy dependence of of charged hadron mid-pseudorapidity density per paticipating quark. Open triangles are wounded quark model results. Solid line is energy dependent functional form of wounded quark model~\cite{[ashwini]}.   }
\label{fig:9} 
\end{figure}
\subsection{\bf{$N_{part}$ and $N_q$ scaling observed in charged hadron multiplicity }}
The energy dependence of the charged hadron pseudorapidity density at mid rapidity normalized by the number of participant pairs for most central collision events is shown in Fig.10 for colliding systems of varying sizes. 
Energy dependence of normalized pseudo-rapidity is found to follow a logarithmic trend for A-A collisions  upto the highest RHIC energy of 200 GeV, which can be well explained as~\cite{[alver]}:
\begin{equation}
\frac{{dn_{ch}}/{d\eta}}{{\langle N_{part}\rangle}/{2}}= 0.78 ln(\sqrt{s_{NN}})-0.4
\end{equation}
However, the measurements in Pb-Pb collisions at 2.76 TeV observed at LHC exceed the above logarithmic dependence.
This logarithmic behaviour describe Au-Au and Pb-Pb data quite successfully with a little disagreement for Cu-Cu data. A better description of the data for energy dependence in A-A collision is given by the following relation~\cite{[alver],[etal]}:
\begin{equation}
\frac{{dn_{ch}}/{d\eta}}{{\langle N_{part}/\rangle}{2}}= 0.77 (s_{NN})^{0.15},
\end{equation}

which is quite successful in describing the data at RHIC and LHC, but fails to explain the lower energy data as it overestimates in lower energy region below 17.3 GeV~\cite{[alver]}. SPM results are shown for $Au-Au$ and $Cu-Cu$ with dashed and dotted lines respectively. CGC and KLN model results are shown by dashed-dotted and dotted lines with error bands~\cite{[rezaeian]}.

In Fig. 11, $N_q$ scaling behaviour of charged hadron pseudo-rapidity density at mid-rapidity is studied in wounded quark scenario over the entire range of energy starting from the lowest AGS energies $\sqrt{s}$=2.4 $GeV$ to the largest LHC energy $\sqrt{s}$=2.76 $TeV$. Solid points in figure are the experimental pseudo-rapidity density data at mid-rapidity normalized by the participating quarks calculated in wounded quark model. Open points are wounded quark model result which is well described by the following functional form:
\begin{equation}
(\frac{dn_{ch}}{d\eta})_{\eta=0}/{N_q}= 0.37 + 0.12 ln^{2}{(\sqrt{s_{NN}})}
\end{equation}
A little deviation from this functional from can be clearly seen for $Cu-Cu$ collisions at 19.6 and 62.4 $GeV$, which arises due to the difference in number of quark collisions  from $Au-Au$ collisions at the same energy and signifies the role of hard processes involved.

\subsection{\bf{Scaled pseudo-rapidity density in A-A collisions as a function of centrality}}
\begin{figure}
\begin{center}
\includegraphics[width=15 cm]{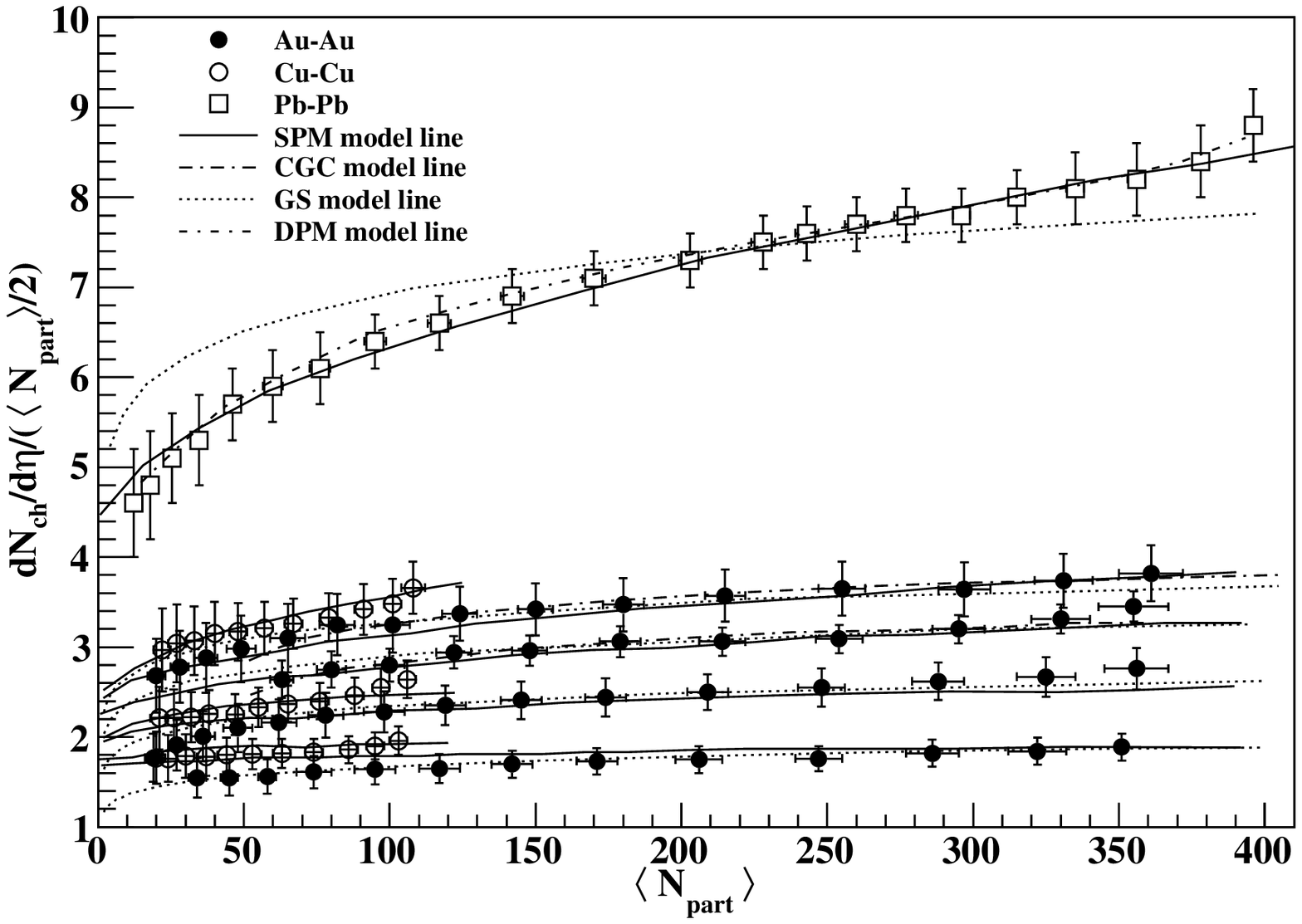}
\end{center}
\caption{Centrality dependence of scaled mid-pseudorapidity density of charged hadrons with different colliding nuclei at RHIC and LHC energies~\cite{[alver],[g.aad]} . Solid lines are the SPM results~\cite{[bautista]}  dash-dotted lines are CGG model results and dotted lines are GS results~\cite{[armesto],[albacete],[gs]}.  }
\label{fig:5} 
\end{figure}

In heavy-ion collisions, nuclei being extended objects collide at various impact parameters depending on the degree or extent of overlap region of the two colliding nuclei. This degree or extent of overlap region formed during the collision is generally referred to as the degree of centrality of the interaction event.  It serves as an appropriate tool to make the suitable comparison between the measurements performed in collider experiments and the theoretical calculations available till date. The mid-rapidity density of charged hadrons normalized to the number of participant pairs (${\langle N_{part}\rangle}/2$) with a variety of colliding systems ($Au-Au, Cu-Cu$ $\&$ $Pb-Pb$) at different energies as a function of centrality (${\langle N_{part}\rangle}$) are shown in Fig. 12. Geometrical scaling model of N. Armesto et al.~\cite{[armesto]} with a strong dependence of the saturation scale on the nuclear mass and collision energy explains the Au-Au multiplicity data quite well and establishes a factorization of energy and centrality dependences in agreement with the data but at LHC energy (2.76 TeV) it predicts a rather weak variation with centrality. On the other hand, the String Percolation model (SPM)~\cite{[deus],[bautista]} driven by the same power law behaviour in both $p-p$ and $A-A$ collisions provide a suitable description of the existing multiplicity data at both RHIC (for $Cu-Cu ~\& Au-Au$) and LHC (for $Pb-Pb$) energies in a consisitent manner. CGC model results at RHIC energies ($130 ~\& 200 GeV$) and LHC energy ($2.76$ TeV) are shown in Fig. 12 by dashed-dotted line. Geometrical scaling (GS) model results~\cite{[armesto],[albacete],[gs]} at RHIC energies  and LHC energy ($2.76$ TeV) are shown in Fig. 12 by dotted line.

\section{Scaling features of charged hadron mulitiplicity in the high energy collision}
Scaling laws in high energy collisions are of great importance as it describes the case where a physical quantity (observable) becomes solely dependent upon a combination of certain physical parameters. Such a scaling law is often useful due to the fact that it can provide us enough information about the underlying dynamics of the observed phenomenon in such collision processes. Whereas, the violation of the scaling law for a certain value of the scaling parameters will be a strong indicator of new physical phenomenon occuring in these collisions.
   
\subsection{\bf{KNO Scaling}}
The energy dependence of the multiplicity distribution observed in high energy collision using a variety of colliding systems is one of the primary issue in search for any systematics/scaling in multiparticle production. Based on the assumption of the Feynman scaling for the inclusive particle production cross section at asymptotic energies Koba, Nielsen and Olesen (KNO) in 1972 proposed scaling property for the multiplicity distribution at asymptotic energies~\cite{[kno]}. According to which the normalized multiplicity distributions of charged particles  should become independent at asymptotic energies. According to KNO scaling, the probability P(n) to produce n charged particles in the final states is related to a scaling function $\psi (z)$ as follows~\cite{[kno]}:
\begin{equation}
\psi (z) =  {\langle n \rangle} P(n) =  {\langle n \rangle} \frac{\sigma_n}{\sigma_{tot}},
\end{equation} 
where $\psi (z)$ is universal function independent of energy  and the variable $z = \frac{n}{\langle n \rangle}$ stands for normalized multiplicity. $\sigma_n$ and  $\sigma_{tot}$ corresponds to n particle production and the total cross section respectively. Thus, the rescaling of $P_n$ measured at different energies via stretching or shrinking the axes by the average multiplicity $\langle n \rangle$ leads the rescaled curves coinciding with one another. However, the experimental data show only an approximate KNO scaling behaviour for multiplicity istributions in the different energy regions and hence Buras et. al~\cite{[buras]} proposed Modified KNO scaling in which variable $z' = \frac {n-\alpha}{\langle n- \alpha \rangle}$. Here parameter $\alpha$ depends solely on the reaction. In order to describe the properties of multiplicity distributions of final state particle at different energies, it is more convenient to study their moments. KNO scaling also implies that the multiplicity moments 
\begin{equation}
C_q = \frac{\sum_n n^q P_n}{(\sum_n n P_n)^q}=\frac {\langle n^q \rangle}{{\langle n \rangle}^q} \nonumber
\end{equation}
 are energy independent since
\begin{equation}
C_q = {\langle Z^q \rangle} = \int Z^q \psi (z)dZ,
\end{equation} 
where q is the order of the moment.
\begin{figure}[htbp]
  \begin{minipage}[b]{0.5\linewidth}
    \centering
    \includegraphics[width=\linewidth]{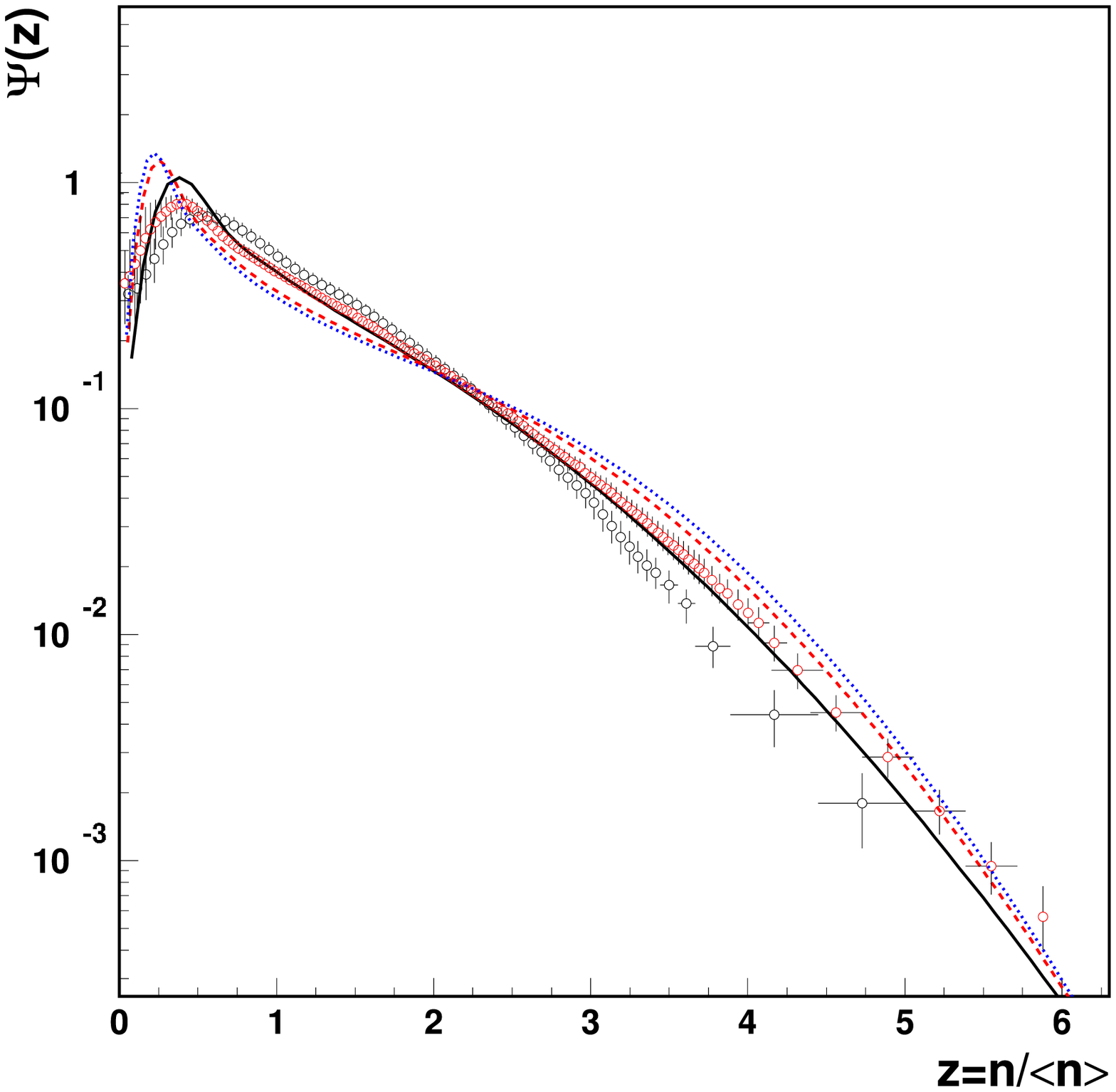}
    \caption{The charged hadron multiplicity distributions in KNO form at $\sqrt{s_{NN}}$ = 0.9 TeV and 7 TeV in pseudorapidity interval, $|\eta|<2.4$ along with comparison to the experimental data. Figure is taken from Ref.~\cite{[capella2013]}.}
    \label{fig:13}
  \end{minipage}
  \hspace{0.5cm}
  \begin{minipage}[b]{0.5\linewidth}
    \centering
    \includegraphics[width=\linewidth]{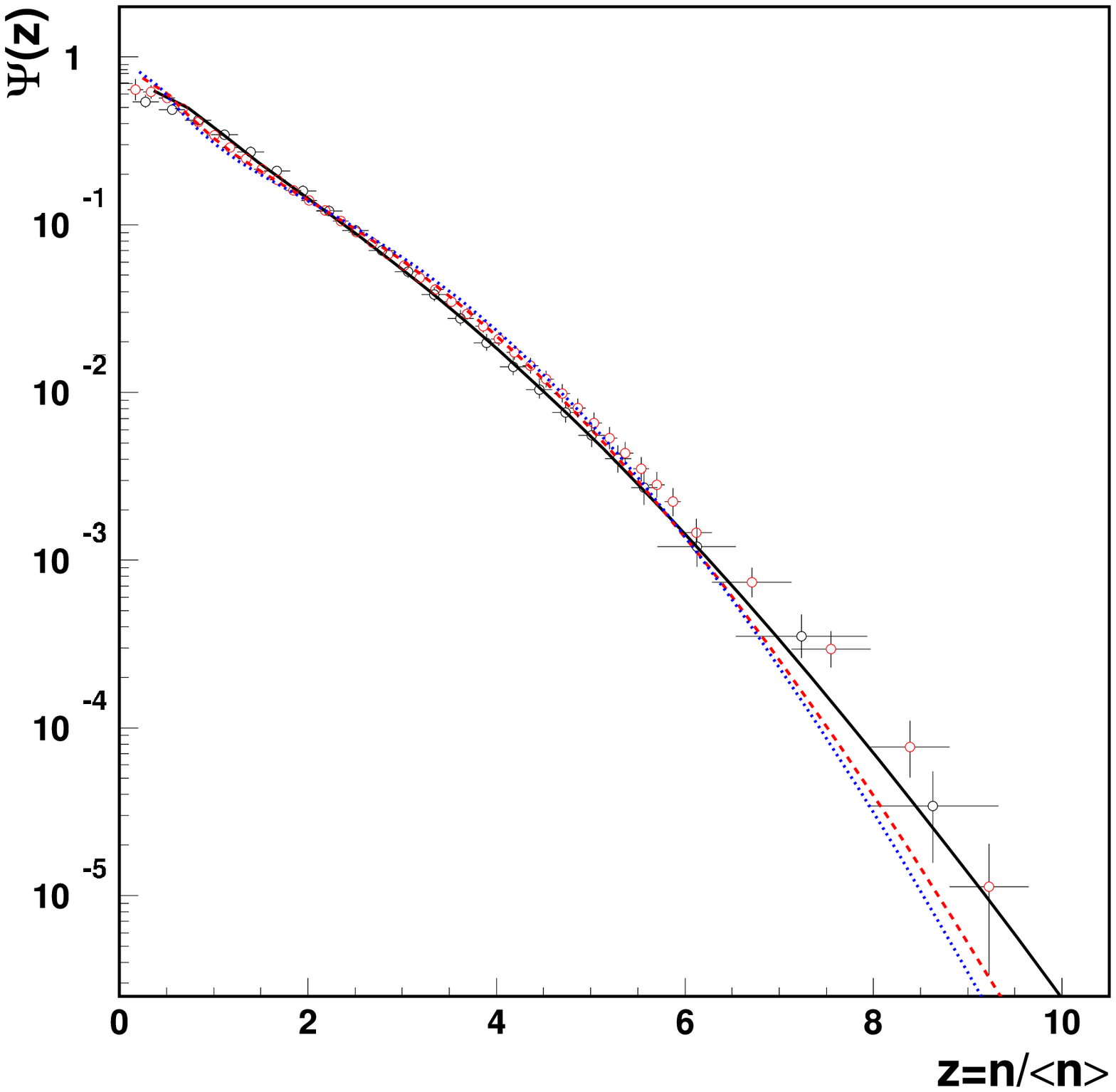}
    \caption{The charged hadron multiplicity distributions in KNO form at $\sqrt{s_{NN}}$ = 0.9 TeV and 7 TeV in pseudorapidity interval, $|\eta|<0.5$ along with comparison to the experimental data. Figure is taken from Ref.~\cite{[capella2013]}.}
    \label{fig:14}
  \end{minipage}
\end{figure}

Indeed, KNO scaling was found to be roughly valid upto the highest ISR energies~\cite{[thome],[breakstone]} but the violation of KNO scaling was first observed to the UA5 data of $p-\bar{p}$ collisions at $\sqrt{s}$=546 GeV~\cite{[violation]} at the CERN collider which provoked a great deal of discussion among the physics community. A high multiplicity tail and a change of slope~\cite{[alexopoulos],[ansorge]} observed in the distribution was interpreted as evidence for multi-component structure of the final state~\cite{[matinyan],[giovannini]}. Later on, a comparative study of charged particle multiplicity distributions~\cite{[alexopoulos]} arising from non single diffractive inelastic hadronic collisions at $\sqrt{s}$=30 GeV to 1800 GeV including the data in UA5(SPS) and E735 (Tevatron) experiments again confirmed the deviations from so called KNO scaling behaviour in the form of a shoulder like structure clearly appeared in the collider data. This shoulder like structure observed in the collider data arises due to the superposition of the distribution of the particles from some other process different from KNO producing process incoherently superimposed on the top of the KNO producing process~\cite{[giovannini1]}. A strong linear increase of the $C_q$ moments with energy and strong KNO scaling violation at $\sqrt{s}$=7 TeV in form of the observed change of slope in $P_n$ confirm these earlier measurements.  Recently, A. Capella et al.~\cite{[capella2013]} have computed pp multiplicity distributions at LHC in the framework of a multiple-scattering model(DPM) as shown in Fig.13 and Fig. 14. Multiple-scattering models do not obey KNO scaling. Indeed, the multiple scattering contributions, which give rise to long-range rapidity correlations, become increasingly important when $s$ increases and since they contribute mostly to high multiplicities they lead to KNO multiplicity distributions that get broader with increasing $s$. On the other hand, the Poisson distributions in the individual scatterings lead to short-range rapidity correlations and give rise to KNO multiplicity distributions that get narrower with incresing $s$. Due to the interplay of these two components the energy dependence of the KNO multiplicity distributions (or of its normalized moments) depends crucially on the size of the rapidity interval~\cite{[capellaproc]}. For large rapidity intervals the multiple-scattering effect dominates and KNO multiplicity distributions get broader with increasing $s$. For small intervals the effect of the short-range component increases leading to approximate KNO scaling, up to $z\sim 6$. We have shown that the above features are maintained up to the highest LHC energy and that for a given pseudo-rapidity interval ($\eta_0 = 2.4$) the rise of the KNO tail starts at a value of $z$ that increases with energy.

All these observations and sizable growth of the measured non-diffractive inelastic cross-sections with increasing energy clearly indicates towards the importance of multiple hard, semi-hard and soft partonic subprocesses in high energy inelastic hadronic collisions~\cite{[capella],[capella1],[polyakov],[aurenche]}.


\subsection{\bf{Intermittency}}
A remarkably intense experimental and theoretical activity has been performed in search of scale invariance and fractality in soft hadron production processes. In this search, a lot of efforts have been performed by investigating all types of reactions ranging from $e^{+}e^{-}$ annihilation to nucleus-nucleus collisions at different energies. In this investigation, Self-similarity (or intermittency)  is a power-law behaviour of a function of the form $f(x)\sim x^{a}$ and thus it also gives the scaling property $f(\lambda x) \sim {\lambda^{a}}f(x)$. Since the scaling is the underlying property of critical phenomena, it is natural to think of a phase transition wherever scaling (or self-similarity) is found. Self-similarity is also connected with a fractal pattern of the structure. Bialas and Peschanski~\cite{[intermittency],[intermittency1]} introduced a formalism to study non-statistical fluctuations as the function of the size of rapidity interval by using the normalized factorial moments of order q. It is expected that scale invariance or fractality would manifest itself in power-law behaviour for scaled factorial moments of the multiplicity distribution in such domain. Scaled factorial moment is defined as:
\begin{equation}
\langle n(n-1).......(n-q+1)\rangle = \sum_{n=1}^{\infty}n(n-1)..........(n-q+1) P_n, \nonumber
\end{equation}
Similarly the scaled factorial moments are defined as:
\begin{equation}
F_q = \langle n(n-1).......(n-q+1)\rangle/ {\langle n\rangle}^q,
\end{equation}
where ${\langle n\rangle}^q = \sum_{n=1}^{\infty} n^q P_n$.

Intermittency is defined as the scale invariance of factorial moments with respect to changes in the size of phase-space cells or bins (say $\delta y$), for small enough $\delta y$:
\begin{equation}
F_{q}(\delta y) \propto {\delta y}^{-{\phi_q}}
\end{equation}
or
\begin{equation}
ln{F_{q}(\delta y)} \sim {-{\phi_q}}ln{\delta y}
\end{equation} 
This give the wide, non-statistical fluctuations of the unaveraged rapidity distribution at all scales. Such a kind of behaviour is a manifestation of the scale invariance of the physical process involved. In Fig. 15 and 16, the intermittent behaviour of charged hadrons produced in the nucleus-nucleus interactions are shown in rapidity- and azimuthal-space~\cite{[ashwini1]}. The slope ($\phi_q$)in a plot of $ln{F_{q}}$ vs -$ln{\delta y}$ at a given positive integer q is called intermittency index or intermittency slope which provide us an opportunity to characterize the apparently irregular fluctuations of the particle density. The term intermittency was chosen in analogy to the intermittent temporal and spatial fluctuations
\begin{figure}[htbp]
  \begin{minipage}[b]{0.5\linewidth}
    \centering
    \includegraphics[width=\linewidth]{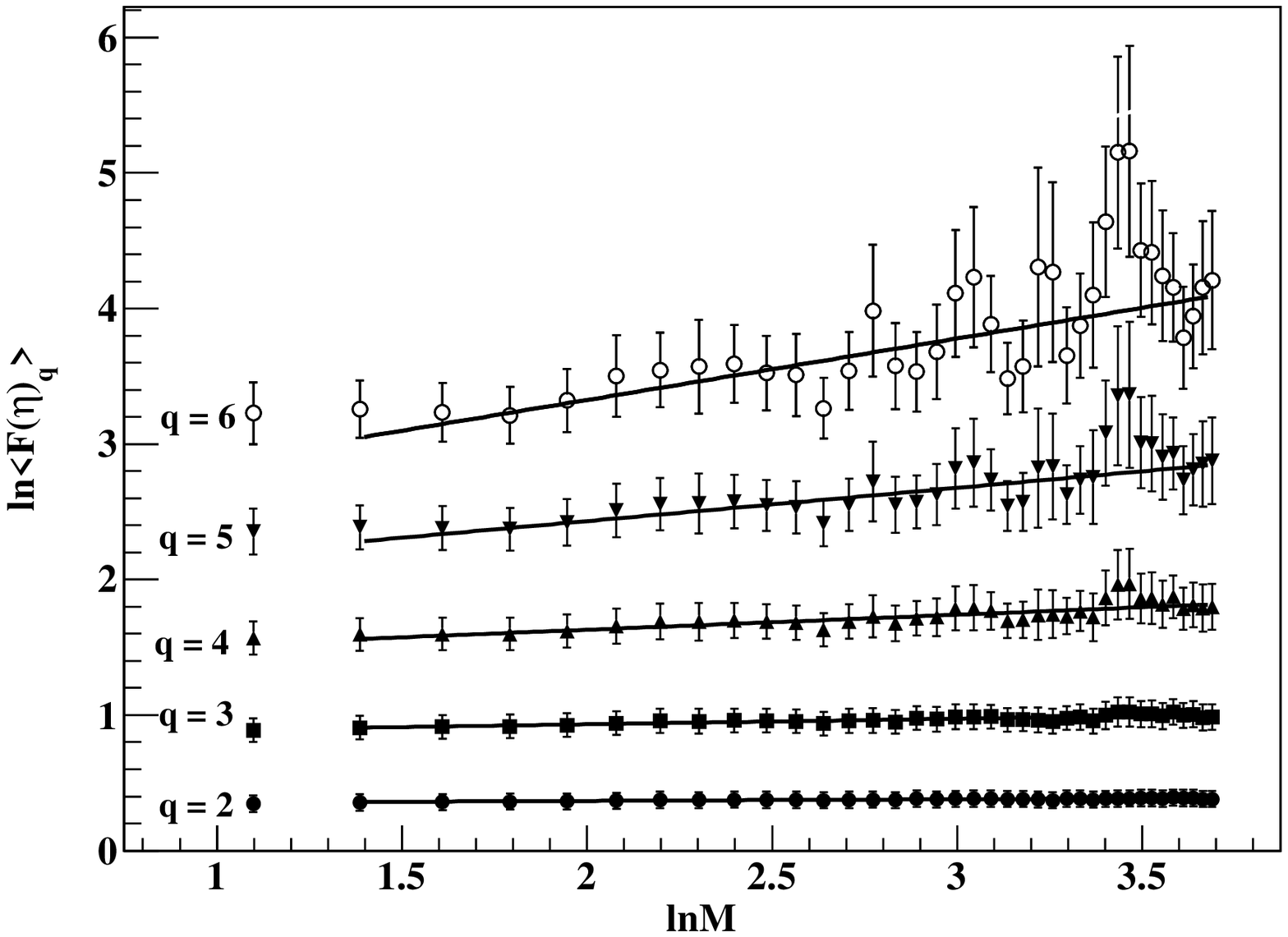}
    \caption{Intermittent behaviour of charged hadrons  produced in $\eta$-space in $^{28}$Si-Ag/Br interactions at 14.6 A GeV.  Figure is taken from Ref.~\cite{[ashwini1]}.}
    \label{fig:15}
  \end{minipage}
  \hspace{0.5cm}
  \begin{minipage}[b]{0.5\linewidth}
    \centering
    \includegraphics[width=\linewidth]{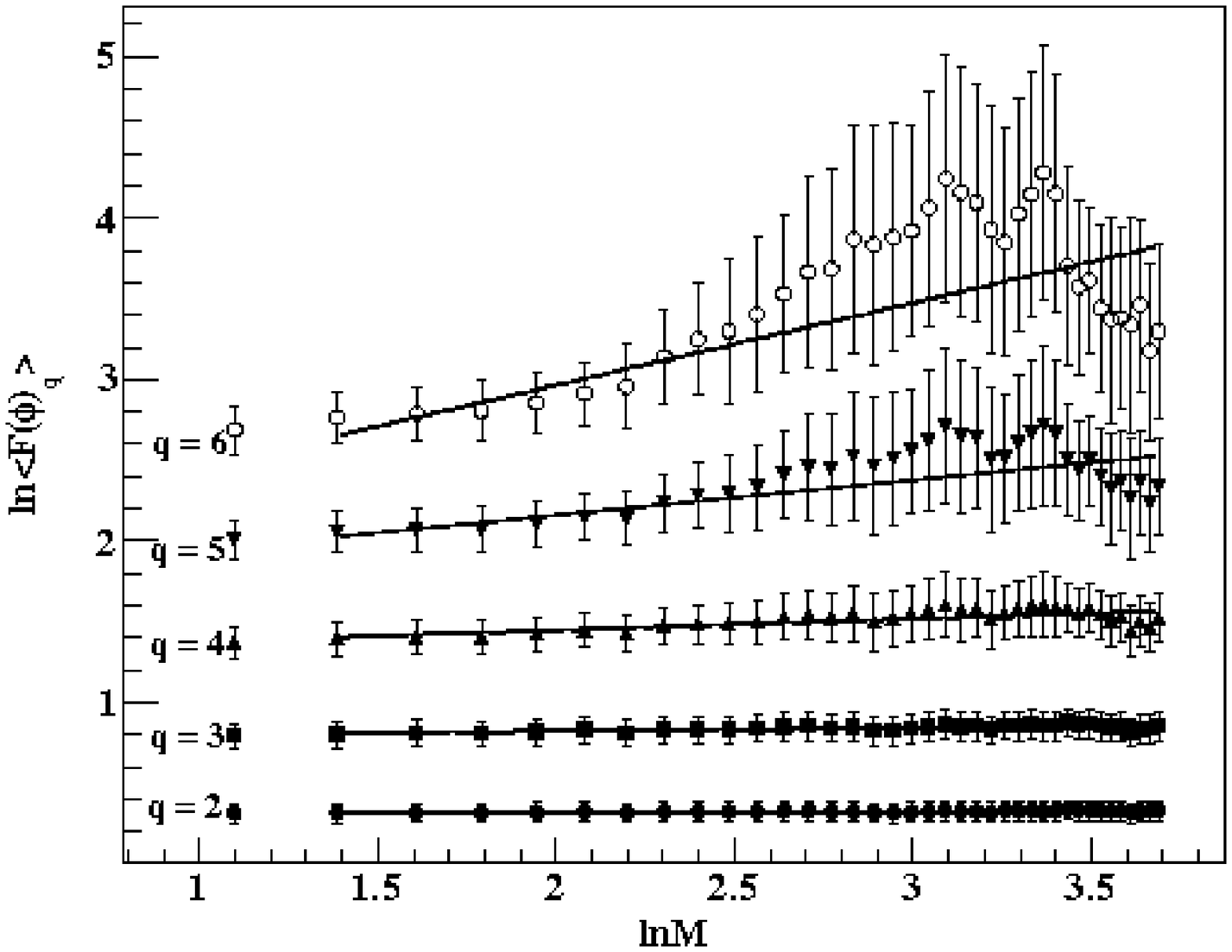}
    \caption{Intermittent behaviour of charged hadrons produced in $\phi$-space in $^{28}$Si-Ag/Br interactions at 14.6 A GeV. Figure is taken from Ref.~\cite{[ashwini1]}. }
    \label{fig:16}
  \end{minipage}
\end{figure}



If there is no dynamical contribution to the multiplicity fluctuation, such as due to a phase transition or any other type of  mechanism, $P_n$ should exhibit the Poisson distribution, reflecting only the statistical fluctuations. The strength of the dynamical fluctuation vary from one event to another event due to the difference in initial condition. In such case, $F_q$ would have no dependence on $\langle n \rangle$ and therefore on phase-space bin($\delta y$). The scaled factorial moments (SFMs) reduce the statistical noise which is present in event with finite multiplicity. Moreover, the method of SFMs is a potentially suitable to investigate the multiparticle correlation on small scales.

A clear signal of the intermittency in $e^{+}e^{-}$ data  was observed for the first time by HRS results~\cite{[hrs],[hrs1]}, shortly followed by TASSO collaboration~\cite{[tasso]} at a center-of-mass energy of about 35 GeV by performing the analysis in 1D (rapidity(y)-space) and 2D (rapidity-azimuthal($y-\phi$) space). At LEP energies, the DELPHI~\cite{[delphi]}, ALEPH~\cite{[aleph]} and OPAL~\cite{[opal]} collaborations performed the intermittency analysis in 1D and 2D distributions. The LUND parton model (JETSET PS) predictions were also found to be consistent with the data. The CELLO collaboration analysed the 3D intermittency signal in $e^{+}e^{-}$ annihilation and  found good agreement with the LUND model.
The hadron-proton ($\pi^{+}-p$ and $K^{+}-p$) collisions were analyzed by the NA22 collaboration. The data for $p-p$ collisions at  630 GeV were analyzed by the UA1 Collaboration and an indication for the increase of the intermittency signal for the low-multiplicity sample was found. A multi chain version of DPM including mini-jet production was compared to NA22 and UA1 data but the slopes were found to be too small~\cite{[bopp],[wolf],[kittel]}. The NA22 group performed the analysis for hadron-nucleus interactions and they found a weaker intermittency signal for large targets. The hadron-nucleus and nucleus-nucleus interactions were also studied by KLM collaboration in 1d and 2d. Decrease in the intermittency signal was observed for larger projectile nuclei and this decrease is smaller than expected from the increase of the mean multiplicity in the collision. The EMU-01 collaboration performed the intermittency analysis for different colliding systems and found a similar dependence. Further, the results for the SFMs in 3D spectra for $e^{+}e^{-}$, hadron-proton, hadron-nucleus and nucleus-nucleus collisions have shown that the dependence of the SFM on the resolution is stronger than a power law.  The failure of Monte Carlo calculations to replicate the observed strength of the intermittency signal is quite evident in the nucleus-nucleus interactions~\cite{[cps]}. A deeper investigations on 1D, 2D and 3D phase space in different kinematic variables ($\eta$, $\phi$, $p_T$) was performed by OPAL collaboration for high statistics study at LEP upto fifth order and down to very small bin sizes~\cite{[abbiendi1],[abbiendi2]}. Later on, it was shown that the Negative Binomial distribution (NBD) faces difficulties to describe the high statistics genuine multiparticle correlations and no one of the conventional multiplicity distributions, including NBD can describe the high statistics data on intermittency of OPAL measurements~\cite{[e.sarkisyan],[kittel1]}. \\
Here, we discuss some important systematics and regularities observed in recent years in this intermittency  study.
The phase-space density of multiparticle production is anisotropic, and the upward bending in the SFM plot is a direct consequence of this anisotropy~\cite{[van.hove],[van.hove1]}. Thus, it is suggested that in the (2d) SFM analysis, the phase-space should be partitioned asymmetrically~\cite{[l.liu]} for taking into the account the anisotropy of the phase-space by introducing a 'roughness' parameter called as the  Hurst exponent ($H$). The Hurst exponent is calculated by fitting the (1d) SFM with the Och's formula~\cite{[w.ochs],[ochs1],[ochs2]}. For $H < 1.0$, the $\phi$-direction is partitioned into finer intervals than the $\eta$-direction, for  $H > 1.0$, the vice-versa is true while,  $H=1$ means that the phase-space is divided similarly in both the directions. 
Ochs and Wosiek~\cite{[ochs]} observed that 1D factorial moments even if they do not strictly obey the intermittency power law over the full rapidity range, still obey the generalized power law
\begin{equation}
F_q = c_q [g(\delta y)]^{\phi_q},
\end{equation}
where the function $g$ depends on the energy and the bin width $\delta y$. Eliminating g yields the linear relation
\begin{equation}
ln{F_q} = \bar{c}_q + (\phi_q/\phi_2)lnF_2.
\end{equation}
This relation is found to be well satisfied by the experimental data for various reactions. The dependence of $\phi_q$ on q can be examined by establishing a connection between intermittency and multifractality~\cite{[hwa1]}. The generalized Renyi dimensions ($D_q$) are related to $\phi_q$ as
\begin{equation}
D_q = D-\frac{\phi_q}{q-1}, 
\end{equation}
where D being the topological dimension  of the supported space and the anamolous dimensions $d_q$ are defined as 
\begin{equation}
d_q = D-D_q =\frac{\phi_q}{q-1}. 
\end{equation}
If we plot the ratio of anamolous dimensions 
\begin{equation}
\frac{d_q}{d_2} = \frac{\phi_q}{(q-1)\phi_2}, 
\end{equation}
 as a function of the order q for various reactions, the q-dependence is claimed to be indicative of the mechanism causing intermittent behaviour and all points fall on a universal curve parameterized as
\begin{equation}
d_q/d_2 = \phi_q/(q-1)\phi_2 = (q^{\mu}-q)/(2^{\mu}-2). 
\end{equation}
 Above parameterization has been derived by Brax and Peschanski~\cite{[brax]} for a self-similar parton-branching process. $\mu$ ($0 \le \mu \le 2$) is called Levy stable index~\cite{[brax1]}. For $\mu =2$ Eq. (69) reduces to Eq. (68). The multifractal behaviour characterized by Eq. (68) and (69) reduces to a monofractal behaviour~\cite{[satz],[satz1],[a.bialas]}. $\mu = 0$ will imply an order-independent anamolous dimension which will happen if intermittency were due to a second-order phase transition~\cite{[wolf]}. Consequently, monofractal behaviour may be a signal for a quark-gluon plasma phase transition.   A noticeable experimental fact that the factorial moments of different order follow simple relation (65), means that correlation function of different orders are not completely independent but are interconnected in some way.
Intermittency can be further studied in the framework of Ginzberg-Landau (GL) model as used to describe the confinement of magnetic fields into fluxoids in type II superconductor. According to which, the ratio $d_q/d_2$ should respect the relation
\begin{equation}
d_q/d_2 = (q-1)^{\nu-1}, \nu = 1.304,
\end{equation}
with $\nu$ being a universal quantity independent of the underlying dimension. The value of $\nu$ was first derived analytically by R. C. Hwa et al.~\cite{[hwa2]} in Ginzburg-Landau model and was confirmed later in a laser experiment~\cite{[hwa3]}.  
If there is any phase transition, it may not necessarily always be a thermal one as the new phase formed is not essentially characterized by the thermodynamical parameters. The possibility of the simultaneous existence of two nonthermal phases (in analogy to different phases of the spin-glass) can be investigated by the intermittency parameter~\cite{[brax1]}
\begin{equation}
\lambda_q = \frac{\phi_q +1}{q}
\end{equation} 
In case of the existence of such different phases in a self-similar cascade mechanism, $\lambda_q$ should have a minimum at some value $q = q_c$. The regions $q < q_c$ resembles liquid phase with a large nuber of small fluctuations and $q > q_c$ resembles dust phase with a small number of large fluctuations. 
Another way to measure the intermittency strength in terms of the intermittency index in the framework of a random cascading model like the $\alpha$-model~\cite{[lianshou]}. According to which the strength parameter $\alpha_q$ is related to the Renyi dimensions as~\cite{[bershadskii]}
\begin{equation}
\alpha_q = \sqrt{\frac{6ln2}{q}(D-D_q)}
\end{equation} 
A thermodynamic interpretation of multifractality can also be given in terms of a constant specific heat C, provided the transition from monofractal to multifractal is governed by a Bernoulli type of fluctuation. Bershadskii proposed a phenomenological relation among the $D_q$ and C as~\cite{[bershadskii]}
\begin{equation}
D_q = D_{\infty} + \frac{Clnq}{q-1},
\end{equation} 
C =0 in the monofractal phase which becomes non-zero finite in the multifractal phase. 

The experimental data on intermittency do not satisfy the criterion for the phase transition and thus one can say the observed intermittency patterns are not suitable probe for QGP formation~\cite{[cps]}.  
\subsection{\bf{Negative binomial multiplicity distribution}}
\begin{figure}
\resizebox{1.0\textwidth}{!}{%
  \includegraphics{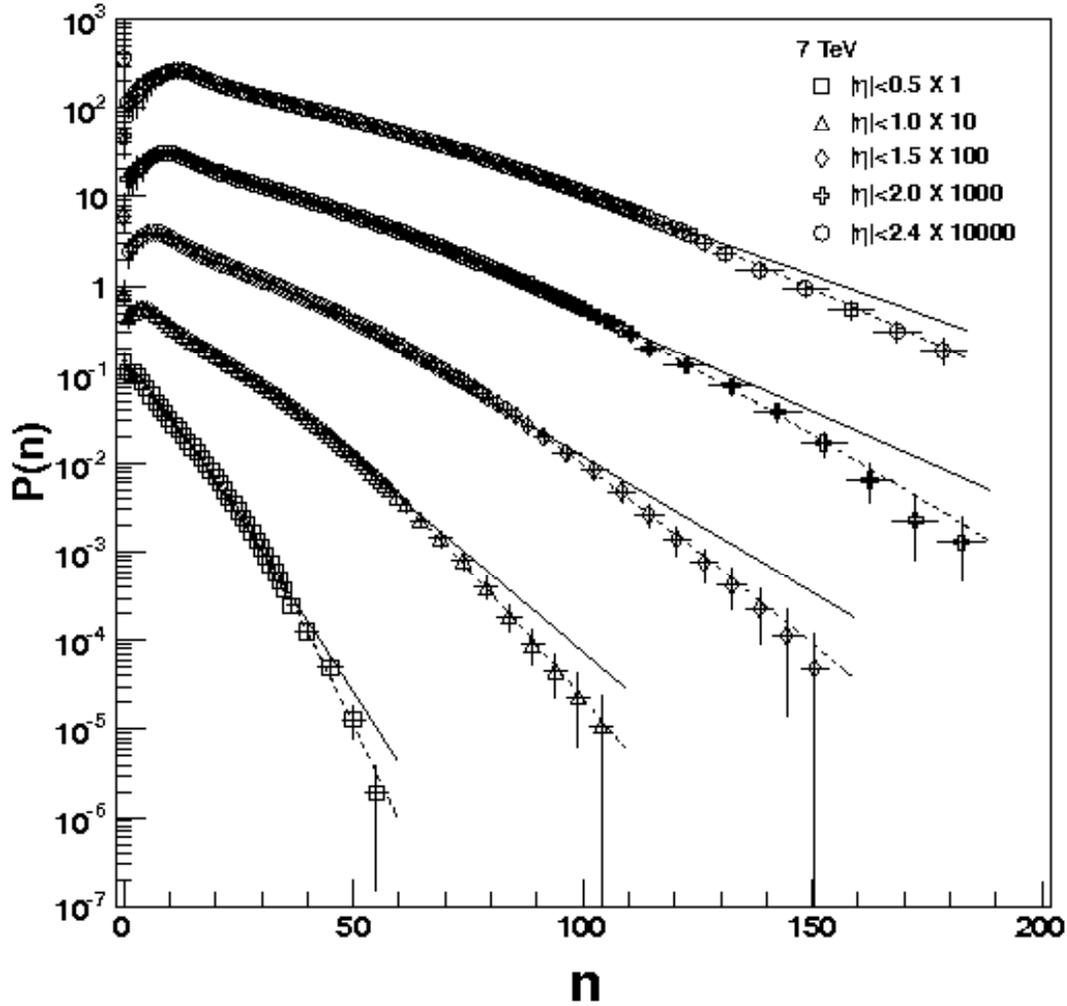}
}
\caption{Primary charged hadron multiplicity distributions for $|\eta|<0.5$ to 2.4 for $\sqrt{s_{NN}}$ = 7 TeV. The solid lines along the data-points correspond to respective single NBD while the dashed lines correspond to respective of Two-NBD. Figure is taken from Ref.~\cite{[premomoy]}.}

\label{fig:1}       
\end{figure}

Another interesting and well known feature observed in multiplicity distribution of produced charged particles in high energy collisions is the occurrence of a negative binomial distribution (NBD) which holds good for a variety of collision processes. It has been observed experimentally for hadron-hadron, hadron-nucleus and nucleus-nucleus collisions over a wide range of energy in finite rapidity as well as in the full phase space. It has further been found that the NBD is valid not only for hadronic, but for semileptonic and leptonic processes as well which emphasize that it is a general property of multiparticle production process regardless of type of colliding particles. To be more specific, the NBD qualitatively describes well the multiplicity distribution almost in all inelastic high energy collision processes except for the data particularly at the highest available collider energies where some deviations seem to be apparent. The negative binomial probability distribution for obtaining n charged particles in the final state is given as follows:
\begin{equation}
P_{n}(n,k) = \left( \begin{array}{c} n+k-1 \\ n  \end{array} \right)\left( \frac{\bar{n}}{k} \right)^n \left( 1+ \frac{\bar{n}}{k} \right)^{-n-k},
\end{equation} 
where $\bar{n}$ and k are two parameters varying with energy and determined by the experimental data. The parameter n has the interpretation of average multiplicity and k is the parameter influencing the width of the distribution . These two parameters are related to the dispersion D = $\sqrt{\bar{n^2}-{\bar{n}}^2}$ as follows:
\begin{equation}
D = \sqrt{\bar{n}+ ({\bar{n}}^2/k)}
\end{equation} 
It is observed that average multiplicity ($\bar{n}$) increases with energy ($\sqrt{s}$) and k decreases with energy. Thus, the negative binomial distribution provides a convenient framework for analyzing the energy variation of the shape of the multiplicity distribution in terms of only two energy dependent parameter (i.e, $\bar{n}$ and $k$). In spite of wide range of the applicability of the NBD in high energy, it is still not a very well understood phenomenon. The parameter $k$ used in the NBD is quite interesting quantity as for increasing $k(k \to \infty)$, the probability distribution ($P_{NBD}$) gets narrower tending to be the Poisson distribution while for the decreasing values of  parameter $k$,  the probability distribution becomes broader and broader than the Poisson distribution. For $k=1$, the $P_{NBD}$ is given by the Bose-Einstein (or geometric distribution). Under the limit of large multiplicity ($n_{ch}\to large$), the $P_{NBD}$ goes over to a gamma distribution. The negative binomial distribution with fixed parameter $k$ exhibits $F$ scaling and asymptotic KNO scaling~\cite{[dremin]}. 

Interpretation of the empirical relation (73) in terms of underlying production mechanism common to hadronic, leptonic and semi leptonic processes is still a challenging problem till date. Some attempts have already been made to derive the NBD from general principles using a stochastic model~\cite{[carruthers]}, a string model~\cite{[werner]}, a cluster model~\cite{[hove]}, a stationary branching process~\cite{[chliapnikov]}, and a two-step model of binomial cluster production and decay~\cite{[lso]}. However, it is still difficult to understand why the same distribution fits such diverse reactions. Moreover, we are still lacking a microscopic model explaining the behaviour of parameter k of the NBD distribution which according to fits decreases with the energy. Recently the NBD form has theoretically been derived in a simplified description of the QCD parton shower with or without hadronization.

In particular, the NBD qualitatively describes the multiplicity distributions in almost all inelastic high energy collisions except for the data recently in collider experiment at much higher energies showing the  deviation from the NBD distribution. For the charged particle multiplicity distribution at $\sqrt{s}$=900 GeV SPS energy, a single NBD function could suitably describe the data only for small pseudo-rapidity intervals at the mid-rapidity region whereas for large intervals, shoulder-like structure appeared in the multiplicity distribution. Appearance of sub-structures in multiplicity distributions at higher energies and in larger pseudo-rapidity intervals has been attributed to weighted superposition or convolution of more than one functions due to the contribution of more than one source or process of particle production. The two-component model of A. Giovannini and R. Ugoccioni~\cite{[giovannini1],[giovannini2]} is quite useful in explaining the multiplicity distribution data at the cost of increased number of adjustable parameters, in which they used weighted superposition of two NBDs representing two class of events, one arising from semi-hard events with minijets or jets and another arising from soft-events without minijets or jets. 
\begin{equation}
P_{n}(\sqrt{s},\eta_{c})=\alpha_{soft}(\sqrt{s})P_{n}[{\langle n \rangle}_{soft}(\sqrt{s},\eta_{c}),k_{soft}(\sqrt{s},\eta_c)]+\pagebreak [1-\alpha_{soft}(\sqrt{s})]P_{n}[{\langle n \rangle}_{semihard}(\sqrt{s},\eta_{c}),k_{semihard}(\sqrt{s},\eta_{c})],\nonumber
\end{equation} 
where $\alpha_{soft}$ signifies the contribution of soft events and is a function of $\sqrt{s}$, the other parameters are functions of both the $\sqrt{s}$ and the $\eta_c $, have their usual meanings as in Eq.(74) indicating respective components.
For small pseudo-rapidity intervals, a single NBD functions explains the distribution data at $\sqrt{s}$= 0.9 and 2.36 TeV reasonably well but at $\sqrt{s}$= 7 TeV, a single NBD function appears inadequate while weighted superposition of two NBDs explain the data satisfactorily as shown in Fig. 17. Whereas, for large pseudo-rapidity intervals ($\eta_c < 2.4$), the weighted superposition of two NBDs agree better than a single NBD function with the distribution data for all available LHC energies clearly indicating the role of soft and semi-hard processes in TeV range. 

\subsection{\bf{Limiting Fragmentation or Longitudinal Scaling observed for charged hadrons}}

\begin{figure}
\begin{center}
\includegraphics[width=15 cm]{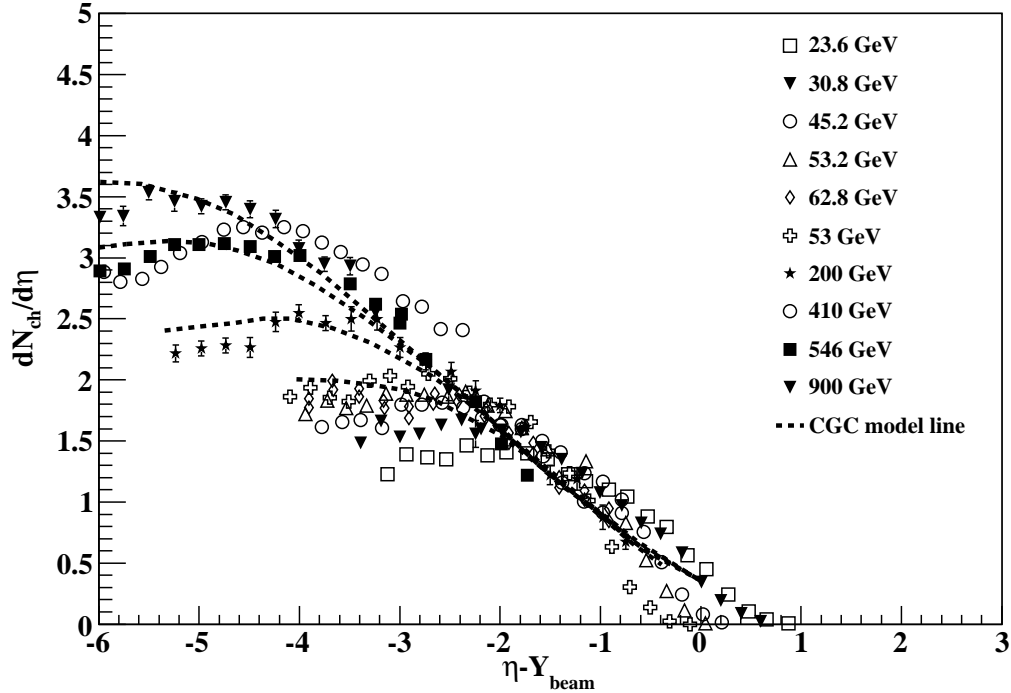}
\end{center}
\caption{Pseudorapidity density of charged hadrons in p-p collisions at different energies as a function of shifted pseudo-rapidity. Data is taken from Ref.~\cite{[b.back],[alver]}.  }
\label{fig:3} 
\end{figure}

\begin{figure}
\begin{center}
\includegraphics[width=15 cm]{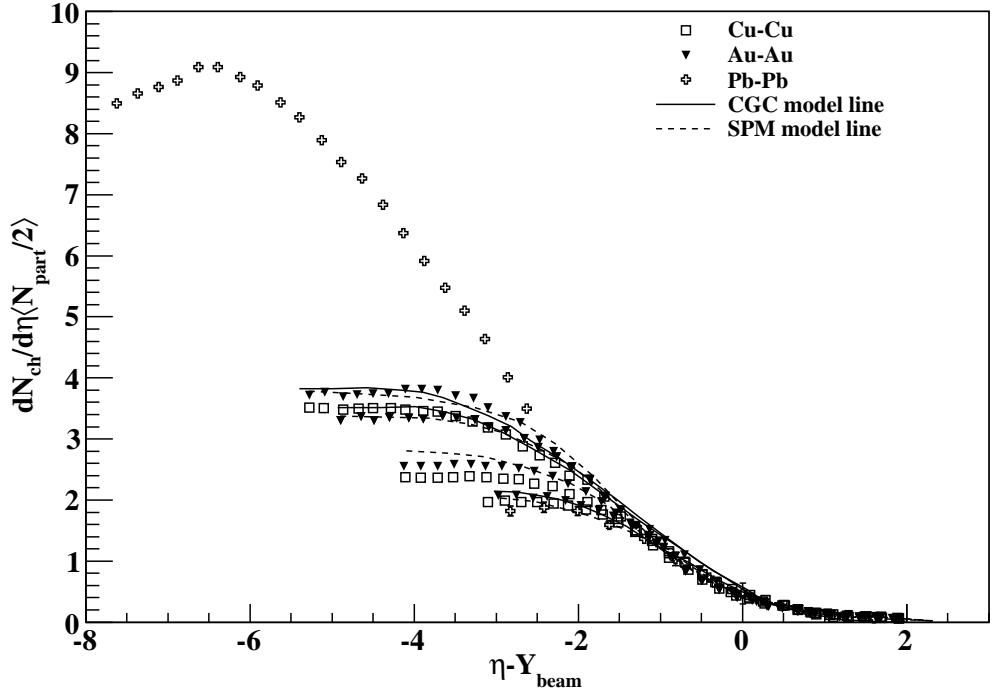}
\end{center}
\caption{Charged hadron pseudo-rapidity density per participant pair as a function of $\eta - Y_{beam}$  at RHIC and LHC energies for different colliding systems. Data are taken from Ref.~\cite{[backcu-cu],[gulbrandsen]}}
\label{fig:1} 
\end{figure}
In order to understand the collision dynamics completely, the study of particle production away from midrapidity region provides some more insight.  The properties of charged hadron production in the fragmentation region can be studied well by viewing the data at different energies in the rest frame of one of the colliding nuclei as the particles near beam and target rapidity are supposed to be governed by the hypothesis of limiting fragmentation~\cite{[benecke]}. The hypothesis of limiting fragmentation for inclusive particle distributions in high energy hadron-hadron collisions was first proposed by J. Benecke et al.~\cite{[benecke]} in 1969. According to this hypothesis, the produced particles in the rest frame of projectile or target will approach a limiting distribution e.g., the pseudo-rapidity density ($\frac{dn}{d\eta}$) as function of $\eta'$ ($\eta-Y_{beam}$) approach a limiting curve in the fragmentation region. This picture is based on the geometrical picture of scattering as considered by Yang and co-workers~\cite{[yang],[yang1],[yang2]}. According to their assumption, in the laboratory frame the projectile nuclei undergoes a Lorentz contraction in the form of a thin disk during the collision with the target nuclei which gets further and further compressed with the increasing energy. However, the momentum and quantum number transfer process between the projectile and target does not appreciably change during this compression. This behaviour leads to limiting distribution of the produced particles in the fragmentation region independent of the center-of-mass energy. Central to this limiting fragmentation hypothesis was the assumption of the constancy of the total hadronic cross sections  at asymptotically large center-of-mass energies. In other words, the probability of the interaction does not change rapidly with the further increase of energy of the incident colliding nuclei. It is expected that the excitation and breakup of a hadron would become independent of the center-of-mass energy and the produced particles in the fragmentation region would approach a limiting distribution. Later on, it was realized that the total hadronic cross sections are not constant, they grow at a slow rate with the increasing center-of-mass energy ($\sqrt{s}$)~\cite{[donnachie],[block]}. This slow increase in the cross sections is still continued upto the LHC energy.   In spite of this fact limiting fragmentation is found to remain valid over a wide range of energy. 
This asymptotic property has been observed experimentally in a variety of collision processes such as hadron-hadron~\cite{[alner],[jones]}, hadron-nucleus~\cite{[jones],[cherry]}, and nucleus-nucleus~\cite{[bearden],[i.g.bearden],[ito],[b.b.back],[b.back],[back],[back11],[adams],[j.adams],[adamovich],[naghy],[ashwini1],[mohanty]} for produced charged hadrons in fragmentation region at different energies. It can be clearly observed that this scaling feature covers a more extended range of $\eta$ than expected according to the original hypothesis of limiting fragmentation due to which the term extended longitudinal scaling is often used to describe this phenomenon. For most central collision, the distributions in fragmentation region are indeed observed to be independent of the collision energy over a substantial shifted pseudorapidity (i.e., $\eta' = \eta - Y_{beam}$) range. In Fig. 18, a compilation of experimental data to study the limiting fragmentation behaviour in $p-p$ collisions is shown along with the CGC model results~\cite{[f.gelis]}. The CGC calculations in $p-p$ collisions with the parameters $\lambda_0 =0.15$ and $\lambda_s =0.69$ are shown with dashed line  at 53, 200, 546 and 900 GeV. A slight discrepancy between the calculations and the experimental data in the mid-rapidity region is seen which arises mainly due to the violation of $k_{\perp}$ formalism in that regime as $k_{\perp}$ formalism becomes less reliable the further one is from the dilute-dense kinematics of the fragmentation regions~\cite{[i.balitsky],[krasnitz1],[krasnitz2],[krasnitz3]}. In Fig. 19,  the pseudo-rapidity distributions at different energies with different colliding nuclei lie on a common curve in fragmentation region over a broad range of $\eta'$ indicating that limiting fragmentation holds good for A-A collision over a wide range of energy and is independent of the size of the colliding nuclei. This gives a clear indication that this universal curve is an important feature of the overall interaction and not simply a breakup effect~\cite{[b.b.back]}. In the CGC approach, the physical picture behind the limiting fragmentation is based on the black disk limit. According to which, in the rest frame of the target nucleus, the projectile nucleus is highly contracted and due to its large number of gluons looks black to the partons in the target nucleus which interact with the projectile nucleus with unit probability.~\cite{[jamal]}. The CGC model calculations~\cite{[f.gelis]} with parameters $\lambda_0 =0.0$ and $\lambda_s =0.46$  at $\sqrt{s_{NN}}$ = 19.6, 130 and 200 GeV for central $Au-Au$ collisions is shown with solid line. Again, the descrepencies in the CGC calculations and the experimental data is quite evident which arises due to the violation of $k_{\perp}$-factorization and due to which the decrease in the multiplicity is expected in this regime~\cite{[i.balitsky],[krasnitz1],[krasnitz2],[krasnitz3]}. SPM results~\cite{[lim]} are shown by dashed line in the figure. Similarly, the scaled pseudo-rapidity density in non-central collisions at different energy is also found to exhibit the limiting  fragmentation in fragmentation region over a broad range of $\eta'$. Thus, we see that the hypothesis of limiting fragmentation is energy independent for a fixed centrality. Similar to the $p-\bar{p}$ collisions, the  fragmentation region for Au-Au colliding nuclei grows in pseudorapidity extent with the colliding beam energy.  Longitudinal scaling observed in the hadronic collisions shows a strong disagreement with the boost invariant scenario which predict a fixed fragmentation region and a broad central plateau growing with energy. Recently, Bialas and Jezabek~\cite{[jezabek]} argued that some qualitative features of limiting fragmentation in hadronic collisions can be explained in a two-step model involving multiple gluon exchange between partons of the two colliding hadrons. This mechanism provides a natural explanation of the observed rapidity spectra, in particular their linear increase with increasing rapidity distance from the maximal rapidity and the short plateau in the central rapidity region.   

 In order  to understand the nature of  hadronic interactions that led to limiting fragmentation, a new mechanism is required which can give a better understanding of the subject.  Although, the limiting fragmentation behavior in small $x$ region is quite successfully described by Color Glass Condensate (CGC) approach. However, the procedure employed in this approach needs further improvements. One of them is the impact parameter dependence of the unintegrated gluon distribution functions. Whereas, in large $x$ region, the phenomenological extrapolation is used which also need to be better constrained and consistent

\section{Summary and Conclusions}
In this review, we have attempted to draw certain systematics and scaling laws for hadron-hadron, hadron-nucleus and nucleus-nucleus interactions. Any distinct deviation from these relations observed in ultra-relativistic nuclear collisions will be the indicator of a new and exotic phenomenon occuring there. However, there are other effect which causes deviations in the scaling features such as mixing of contributions from soft and hard processes, etc. Present experimental results show that  the scaling features of charged hadrons previously observed at lower energies do not hold good at higher energies. Thus, we are still lacking a scaling law which is universal to all types of reactions and can give some basic understanding of the mechanism involved in the particle production.  The observed deviations from the well established scaling laws are of great interest as these clearly hint towards the increasing role of hard processes contributing in the production of final state particles at higher energies. These hard processes of the primordial stage of the collision contribute significantly to the hadronization process at the late stage of the evolution. The applicability of the two-component approach in nucleon-nucleon interaction picture and in wounded quark picture to evaluate mid-rapidity density of charged hadrons is quite promising as the weighted superposition of hard and soft processes comes into picture and  successfully describe the available experimental data over a large range of energy. Our study of the charged hadron production using phenomenological models  along with the experimental data essentially highlights some kinds of scaling relations for these high energy collision processes.  $N_{part}$ scaling of pseudo-rapidity density at mid-rapidity holds only for limited range of energy, while for the entire range of energy (including LHC) it does not seem to hold good which clearly hints towards the contributions of some other processes.  On the other hand, the wounded quark model  provides a more realistic explanation of charged hadron production in terms of the  basic parton structure involved in these processes. However, the charged hadron pseudo-rapidity density at mid-rapidity in wounded quark model also does not show an exact $N_q$ scaling type behaviour and the deviations are clearly observed which arises due to difference in the number of quark collisions for different colliding nuclei as involved in two-component model approach used for calculating the pseudo-rapidity density. Thus, both in wounded nucleon picture and in wounded quark picture, the participant scaling does not hold good over the entire energy range. However, two-component picture which involves the relative contribution of both the number of participants and the number of collisions provide a suitable description of the available  experimental data in the entire energy range. It clearly indicates the contributions of both the hard and soft processes involved in the charged hadron production.  Further, the deviations from the KNO scaling behaviour of charged hadron multiplicity distribution  gives a clear indication towards the role of multiple hard, semi-hard and soft partonic sub processes involved in the hadronic collisions. Similarly, the failure of single NBD function to describe the experimental data is clearly visible in TeV energy range also. While the success of weighted superposition of two NBDs in explaining the multiplicity distribution, clearly reflect the contribution of more than one source or processes involved in these collisions.  The well-known limiting fragmentation behaviour of charged hadrons in fragmentation region seems to hold good since the scaling is compatible with the experimental results at LHC energy (2.76 $TeV$). This universal limiting fragmentation behaviour appears to be a dominant scaling feature of the pseudo-rapidity distribution of the charged hadrons in high energy collisions.  Another such phenomenon which is universal to all types of reactions is the scaling properties of multiplicity fluctuations (called as intermittency). The picture in this sector  is  still not very clear and complete which requires furthermore scruitiny.

 In conclusion, the multiparticle production in the ultra-relativistic hadron-hadron, hadron-nucleus, and nucleus-nucleus collisions is still a burning topic in high energy physics because it throws light on the production mechanism. Since we still lack explanation of these processes in QCD, we hope that our investigations regarding universal systematics and scaling relation on charged hadron production will certainly lead us to a correct phenomenological model for the production mechanism. In addition to this, recent studies regarding fluctuations and correlations is expected to provide deeper insight into different stages of the system evolution produced in heavy-ion collisions. One of the key observables in fluctuation-correlation studies is elliptic flow ($v_2$). Elliptic flow provides the strength of interactions among the particles at early stages in the heavy-ion collisions. A consistent and unified description of the elliptic flow measured in $Cu-Cu$ and $Au-Au$ collisions at mid-rapidity can be obtained after scaling $v_2$ with the participant nucleon eccentricity ($\epsilon_{part}$) ~\cite{[hofman],[barbara]}. Further, elliptic flow scaled with $\epsilon_{part}$ for the same number of participating nucleons in $Cu-Cu$ and $Au-Au$ system show a scaling pattern upto 3.5 $GeV/c$ in $p_T$ and across  $\pm 5$ units of pseudo-rapidity~\cite{[hofman],[barbara]}. Thus, it will definitely be interesting to investigate these fluctuations and correlations  in future to draw a more consistent and unified picture of particle production and their evolution in heavy-ion collisions.

\end{document}